\documentclass{aastex}
\usepackage[twocolumn]{emulateapj5}

\submitted{ApJ, accepted (12 Jun 2006)}

\def\ltsima{$\; \buildrel < \over \sim \;$}
\def\simlt{\lower.5ex\hbox{\ltsima}} 
\def\gtsima{$\; \buildrel > \over \sim \;$}
\def\simgt{\lower.5ex\hbox{\gtsima}} 
\def\arcsec{\hbox{$^{\prime\prime}$}}
\def\deg{\hbox{$^\circ$}}

\shorttitle{Large-scale Jet in $z$=3.89 Quasar} 
\shortauthors{Cheung, Stawarz, \& Siemiginowska}

\begin{document}

\title{Confronting X-ray Emission Models with the Highest-Redshift 
Kiloparsec-scale Jets: the $z$=3.89 Jet in Quasar 1745+624}

\author{C.~C. Cheung\altaffilmark{1}}
\affil{Kavli Institute for Particle Astrophysics and Cosmology, Stanford 
University, Stanford, CA 94305; teddy3c@stanford.edu}

\altaffiltext{1}{Jansky Postdoctoral Fellow of the National Radio
Astronomy Observatory.}

\author{\L . Stawarz\altaffilmark{2}}
\affil{Landessternwarte Heidelberg, K\"onigstuhl, and
Max-Planck-Institut f\"ur Kernphysik, Saupfercheckweg 1, 69117 Heidelberg,
Germany}

\altaffiltext{2}{Also Astronomiczne, Uniwersytet Jagiello\'nski, ul. Orla
171, 30-244 Krak\'ow, Poland.}

\author{A. Siemiginowska}
\affil{Harvard-Smithsonian Center for Astrophysics, 60 Garden St.,
Cambridge, MA 02138}

\begin{abstract}

A newly identified kiloparsec-scale X-ray jet in the high-redshift
$z$=3.89 quasar 1745+624 is studied with multi-frequency Very Large Array,
Hubble Space Telescope, and {\it Chandra} X-ray imaging data.  This is
only the third large-scale X-ray jet beyond $z>$ 3 known and is further
distinguished as being the most luminous relativistic jet observed at any
redshift, exceeding $10^{45}$ erg/s in both the radio and X-ray bands.
Apart from the jet's extreme redshift, luminosity, and high inferred
equipartition magnetic field (in comparison to local analogues), its basic
properties such as X-ray/radio morphology and radio polarization are
similar to lower-redshift examples.  Its resolved linear structure and the
convex broad-band spectral energy distributions of three distinct knots
are also a common feature among known powerful X-ray jets at
lower-redshift. Relativistically beamed inverse Compton and `non-standard'
synchrotron models have been considered to account for such excess X-ray
emission in other jets; both models are applicable to this high-redshift
example but with differing requirements for the underlying jet physical
properties, such as velocity, energetics, and electron acceleration
processes. One potentially very important distinguishing characteristic
between the two models is their strongly diverging predictions for the
X-ray/radio emission with increasing redshift. This is considered, though
with the limited sample of three $z>$ 3 jets it is apparent that future
studies targeted at very high-redshift jets are required for further
elucidation of this issue. Finally, from the broad-band jet emission we
estimate the jet kinetic power to be no less than $10^{46}$ erg/s, which
is about 10$\%$ of the Eddington luminosity corresponding to this galaxy's
central supermassive black hole mass ${\cal M}_{\rm BH}$ \simgt $10^9 \,
{\cal M}_{\odot}$ estimated here via the virial relation.  The optical
luminosity of the quasar core is about ten times over Eddington, hence the
inferred jet power seems to be much less than that available from mass
accretion. The apparent super-Eddington accretion rate may however suggest
contribution of the unresolved jet emission to the observed optical flux
of the nucleus. 

\end{abstract}


\section{Introduction\label{section-intro}}

\begin{figure*}
\epsscale{0.75}
\plotone{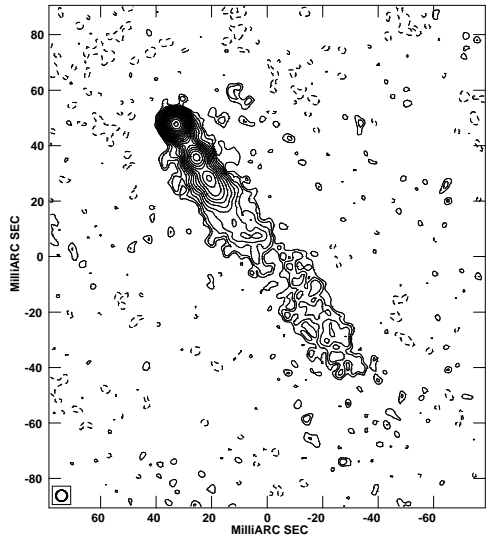}
\figcaption[f1.eps]{\label{figure-1}
VLBI 2.3 GHz map of the parsec-scale jet in quasar 1745+624 from averaging
6 images from the USNO database (4 mas circular beam plotted on bottom 
left).  The quasar core is the peak (239.0 mJy/bm) toward the upper left. The
lowest contour level is 0.3 mJy/bm (2 times the measured rms in the image)
increasing by factors of $\sqrt{2}$.
}
\end{figure*}

\begin{figure*}
\epsscale{1.75}
\plotone{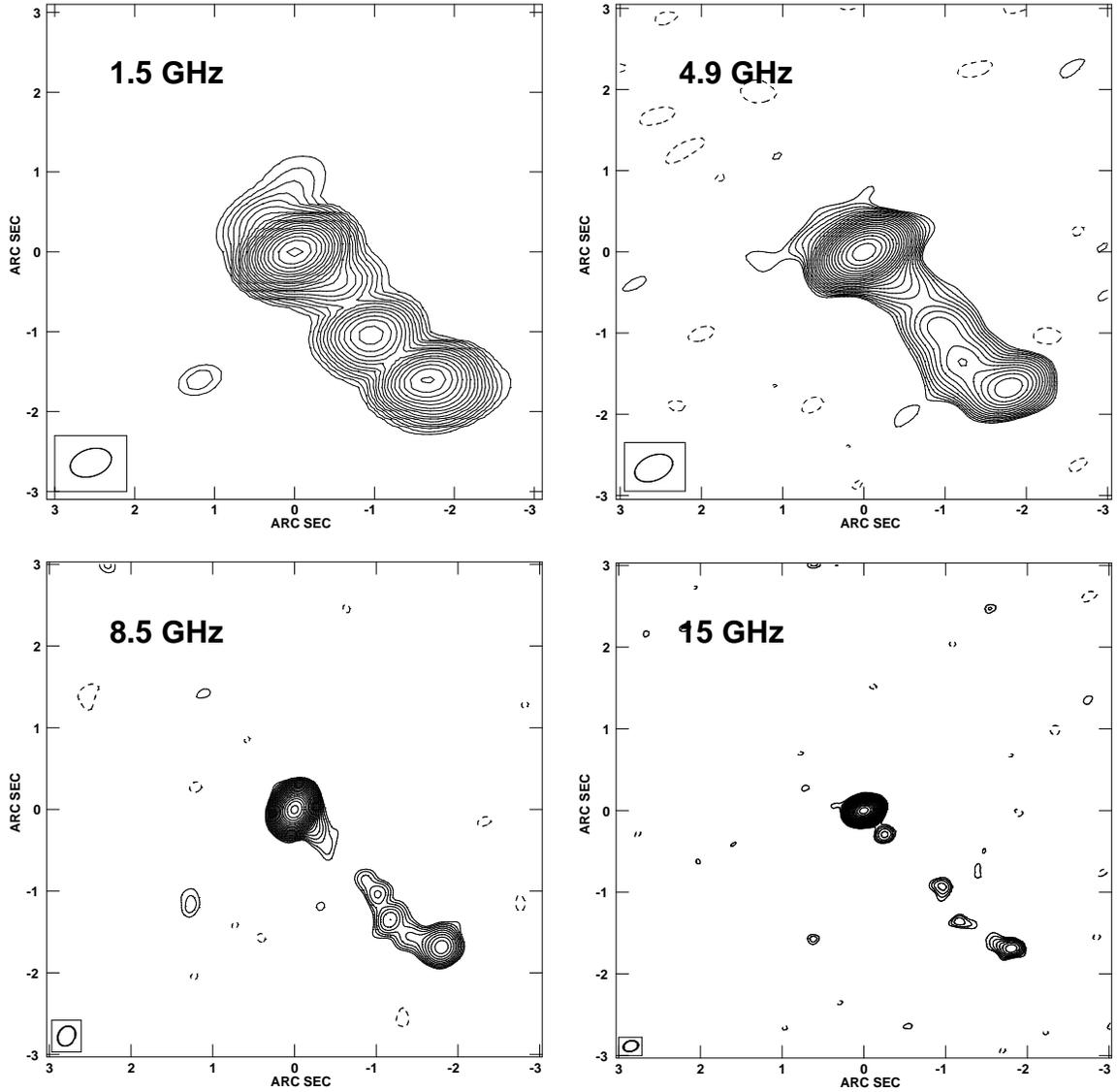}
\figcaption[f2.eps]{\label{figure-2}
Multi-frequency VLA images of 1745+624. The $\sim$2.5$\arcsec$ long radio 
jet extending to the southwest of the core (placed at the origin) is seen
clearly. The lowest contour levels are 0.60, 0.33, 0.40, and 0.45 mJy/beam
(3 times the measured off source rms in the images), at 1.5, 4.9, 8.5, and
15 GHz, respectively. The positive levels (solid contours) are spaced by
factors of $\sqrt{2}$ up to the image peaks of 484, 441, 464, and 577   
mJy/beam. The elliptical restoring (naturally weighted) beams are plotted
at the bottom left corner: their dimensions are
0.530\arcsec$\times$0.340\arcsec\ at PA=--72$^{\circ}$,
0.490\arcsec$\times$0.305\arcsec\ at PA=--67$^{\circ}$,
0.255\arcsec$\times$0.204\arcsec\ at PA=--29$^{\circ}$, and
0.182\arcsec$\times$0.128\arcsec\ at PA=--74$^{\circ}$.}
\end{figure*}

\begin{figure*}
\plotone{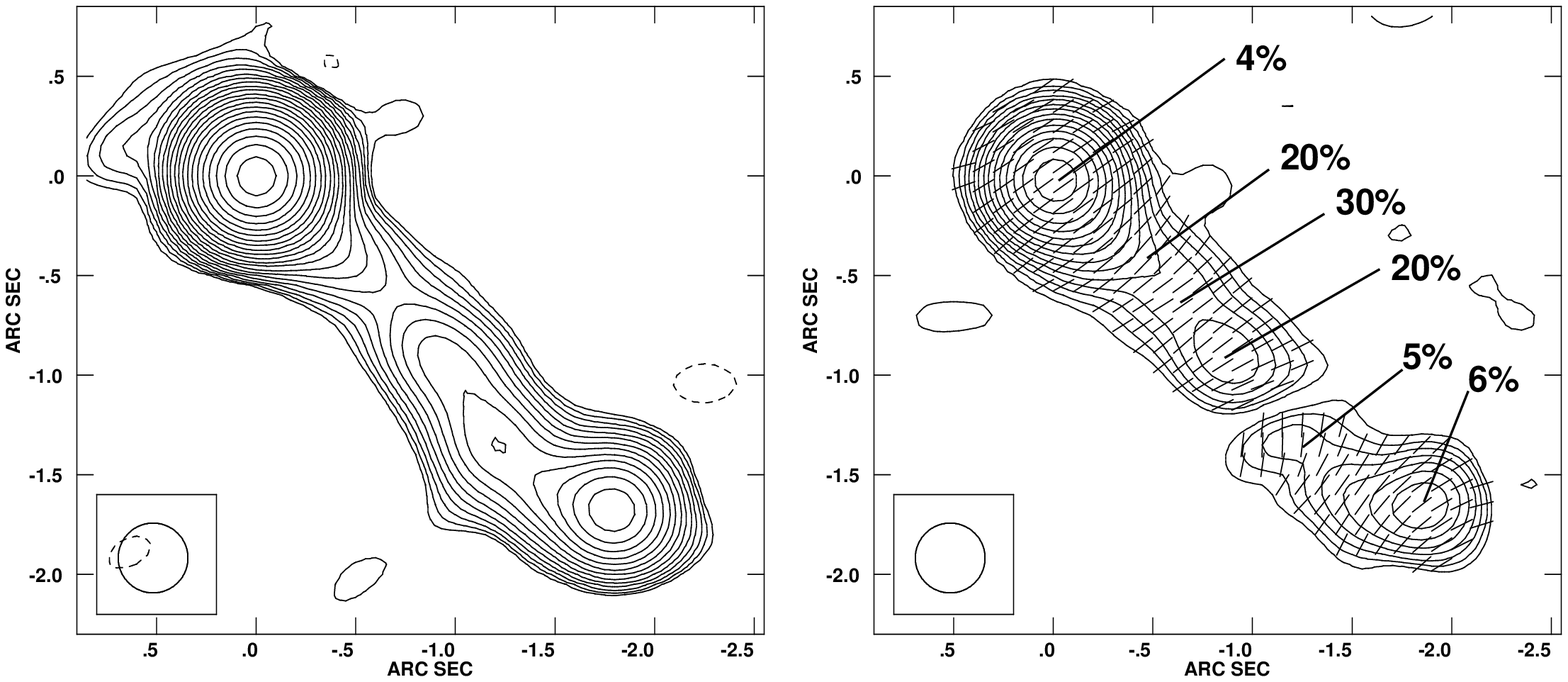}
\figcaption[f3.eps]{\label{figure-3} 
VLA 4.9 GHz total intensity [I; left] and polarized intensity [P; right]
images of 1745+624 at 0.35\arcsec\ resolution (beam plotted in bottom
left). Contour levels begin at 0.35 (I) and 0.16 (P) mJy/bm, and increase
by factors of $\sqrt{2}$ up to peaks of 441.4 (I) and 18.1 (P) mJy/bm. The
tick marks show the orientation of the electric vector position angles,
with a correction of --6\deg\ applied corresponding to the integrated
rotation measure of 28.4$\pm$0.5 radians m$^{-2}$ \citep{ore95}. Selected
fractional polarization levels are indicated.}
\end{figure*}

Quasars form a class of objects that frequently warrant superlatives in
their descriptions:  1745+624 (4C+62.29) is no exception. It was
identified as one of the highest redshift X-ray quasars at the time of its
discovery \citep[$z$=3.87;][]{bec92} from spectroscopic followup of
statistically significant sources from the {\it Einstein} X-ray
Observatory. Further optical spectroscopy by \citet{sti93} refined the
redshift to $z$=3.89, affirmed by \citet{hoo95}; the latter value is
adopted here. After the {\it Einstein} detection, it was established as a
bona fide X-ray source by {\it ROSAT} \citep{fin93}, {\it ASCA}
\citep{kub97}, and {\it BeppoSAX} \citep{don05}. 

By virtue of its high-redshift, it is one of the most radio-luminous
quasars known \citep[cf. Figure~1 of][]{jes03} -- its observed 1.4 GHz
luminosity\footnote{We adopt H$_{0}=71~$km~s$^{-1}$~Mpc$^{-1}$,
$\Omega_{\rm M}=0.27$ and $\Omega_{\rm \Lambda}=0.73$ and have converted
quoted literature values of luminosities and proper motions to this
cosmology. If we assume the cosmology adopted in \citet{jes03}, $L_{\rm
1.4~GHz}=10^{28.4}$ W Hz$^{-1}$ (1 Watt = 10$^{7}$ erg s$^{-1}$).},
$L_{\rm 1.4 GHz}=10^{29}$ W Hz$^{-1}$, is over an order of magnitude
greater than that of the archetype luminous quasar 3C~273 \citep{con93}. 
It is probably no coincidence that its 2.5\arcsec\ long jet \citep[18 kpc
projected;][]{bec92} is also the most luminous radio jet currently known
\citep[that we are aware of; cf. ][]{liu02,che05}, accounting for
$\sim$1/4th of the source's total 1.4 GHz luminosity. 

This quasar is of interest to us because of this prominent radio jet,
visible on both milli-arcsecond
\citep[][Figure~\ref{figure-1}]{tay94,fey00} and arcsecond-scales
(Figures~\ref{figure-2} \& \ref{figure-3}).  We were prompted to identify
such high-redshift systems with large-scale jets after the {\it Chandra}
X-ray Observatory detection of a $\sim$2.5\arcsec\ long X-ray extension
\citep{sie03,yua03,che04} in the $z$=4.3 quasar GB~1508+5714
\citep{hoo95}. Subsequent VLBA imaging of GB~1508+5714 has revealed a
parsec-scale jet that can be traced out to $\sim$100 milli-arcseconds
\citep[0.7 kpc projected;][]{che06}, aimed in the general direction of the
kpc-scale structure, supporting the jet interpretation. These are the two
most distant quasars with a {\it kiloparsec-scale} jet\footnote{Adopting
the conventional requirement of a jet being at least 4 times longer than
it is wide \citep{bri84}. A single feature associated with a jet was
discovered in the more distant $z$=5.47 blazar Q0906+6930 by
\citet{rom04}, but on VLBI scales only.} detected at any wavelength
\citep{che05}. 

The identification of such systems in the early Universe is extremely
interesting for a number of reasons. Jets are signposts for ``active''
black hole/accretion disk systems \citep[e.g.,][]{beg84}, requiring
central nuclei of high-$z$ galaxies to be sufficiently well-developed to
sustain gravitational collapse. At such early epochs, it is remarkable
that the jet production process is apparently efficient and persists long
enough ($>$Myr) to have produced such large (10's--100's kpc-scale)
structures. In this context, jets may even serve as tracers of the
directionality of the BH/disk axis in these early accreting systems, as
chronicled in the ``bumps and wiggles'' in our jet images.  Also, these
jets are an efficient means to shock heat ambient gas thus triggering
early star formation \citep[e.g.,][]{ree89}. 

Among other important issues, such high-redshift systems are potentially
very important to our understanding of the emission processes responsible
for the production of the broad-band jet emission, and the X-rays in
particular. Although the number of quasars with kpc-scale jets detected in
the X-rays is growing \citep{har06}\footnote{See a current on-line census
at:  http://hea-www.harvard.edu/XJET/}, the origin of this radiation is
still in active debate.  \citet{sch02} noted early on that a comparison of
cosmologically distant jets with local examples could in principle provide
crucial arguments to the debate. This is because the two main contending
models for the X-ray emission of large-scale quasar jets -- synchrotron
radiation and inverse Compton scattering of the cosmic microwave
background (IC/CMB) -- have, in their simplest versions, markedly different 
predictions of the X-ray jet properties with redshift (see \S~\ref{section-z}). 

A high-resolution {\it Chandra} image of the $z$=3.89 quasar 1745+624 was
obtained in an early observing cycle (Table~\ref{table-1}) and X-ray
emission from the arcsecond-scale radio jet is in fact detected and is
quite prominent also. Apart from the GB~1508+5714 case mentioned above,
evidence has been presented recently for at least one more high-$z$ quasar
with extended X-ray emission \citep[J2219--2719 at $z$=3.63;][]{lop06}
with a radio counterpart \citep[][and manuscript in preparation]{che05}.
This makes 1745+624 one of only three very high-redshift ($z>$3) systems
with extended X-rays associated with a radio jet. 

With the scarcity of such high-redshift systems and their importance in
current discussions of X-ray jet emission models, we have independently
analyzed the archival {\it Chandra} data. It was also realized that
1745+624 hosts an exceptional jet making it conducive to study. This is
because while both the GB~1508+5714 and J2219--2719 cases each show a
single faint ($\sim$1 mJy at 1.4 GHz) arcsecond-scale feature separated
from their bright nuclei, the 1745+624 jet is 2 orders of magnitude
brighter than the other two cases and shows {\it resolved structure}. In
addition, it is the most luminous X-ray jet observed thus far
\citep[cf.,][]{har06}. 

To fully exploit the potential of the {\it Chandra} observation, we have
also analyzed archival multi-frequency imaging data from the Very Large
Array (VLA), and a Hubble Space Telescope (HST) image to map the spectral
energy distributions of the extended components in the quasar jet. These
data, along with a new deep very long baseline interferometry (VLBI) map
of the extended subarcsecond-scale jet are described in
\S\S~\ref{section-vlbi}--\ref{section-chandra}, with a critical assessment
of the derived spectra of the kpc-scale components in
\S~\ref{section-spectra}. The physical properties of these extended
components are then discussed (\S~\ref{section-discuss}), with comparisons
made to the other two quasars with X-ray detected knots at comparably
high-redshift ($z\simgt$3). These distant examples are compared to other
currently known X-ray jets in quasars at lower redshift ($z\simlt$2) in
light of expectations from synchrotron and inverse Compton emission
models. Our results are summarized in \S~\ref{section-summary}.

\section{Multi-Telescope Archival Data and Analysis\label{section-data}}

\subsection{Deep 2.3 GHz VLBI map of the Parsec-scale Jet\label{section-vlbi}}

VLBI maps of 1745+624 show a prominent parsec-scale jet and the emission
fading with distance out to $\sim$40 mas \citep{tay94,fey00}.  The lower
frequency, 2.3 GHz USNO Radio Reference Frame Image Database maps
\citep[RRFID, ][]{fey00} suggest that the jet is even more extended. We
therefore uniformly reprocessed the six best ($u,v$) datasets at this
frequency from the RRFID (obtained between 1997 March to 1998 August) and
produced images on the same grid, reconvolved with a common 4 mas beam. 
We then averaged the 6 images to suppress the noise in the background
resulting in improved definition of the outer jet (Figure~\ref{figure-1}).
The parsec-scale jet in this ``average" image is now clear out to
$\sim$110 mas (0.8 kpc projected) allowing us to trace
(\S~\ref{section-describe}) the VLBI-scale emission to the outer
structure sampled by the VLA data (Figure~\ref{figure-2} and
\S~\ref{section-radio}). 

At 4 mas resolution, the initial bright segment of the VLBI jet is
oriented at position angle (PA) 212\deg\ and contains two prominent peaks
at roughly 14 mas (22.3 mJy/bm) and 23 mas (10.4 mJy/bm). At $\sim$50 mas
from the core, the jet position angle increases (by $\sim$4\deg) and
transitions to the outer fainter portion.  Curiously, the eastern
ridgeline of the outer portion lines up with the central PA of the inner
jet. The integrated flux of the visible jet in the 2.3 GHz map is
$\sim$110 mJy with the outer portion contributing about 20 mJy of this. 
We measure a jet to counter-jet ratio of $>$50 (defined as the ratio of 
the peak to 3 times the rms noise on the counter-jet side) of the first 
visible peak in the 2.3 GHz map.

\begin{figure*}
\epsscale{1.0}
\plotone{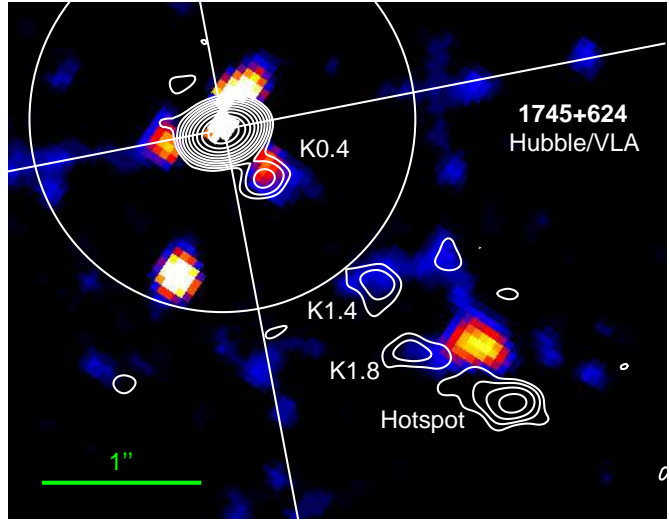}
\figcaption[f4.eps]{\label{figure-4}
HST STIS (color) and VLA 14.9 GHz (contours; see Figure~\ref{figure-2})
images of the immediate field around quasar 1745+624. A model of the
central optical source from fitting a series of elliptical isophotes has 
been subtracted, out to approximately 1.2\arcsec\ (the radius of the
pictured circle). The white lines crossing the nucleus indicate the
directions of the main diffraction spikes in the image. The main knots  
in the jet discussed in the text are labeled.}
\end{figure*}

\subsection{Very Large Array\label{section-radio}}

In light of the high signal-to-noise (S/N) ratio {\it Chandra} detection
of the arcsecond-scale jet in 1745+624 (\S~\ref{section-chandra}), we
desired comparable information about its structure and spectra in the
radio band. 

\citet{bec92} first discovered the extended radio emission in short VLA
observations at 1.5, 5 and 15 GHz. The two main X-ray emitting knots which we
dub K1.4 and K1.8 (indicating their distance from the core in arcseconds),
and the terminal (brightest) jet feature at 2.5\arcsec\ are readily
identifiable in their 5 GHz map.  For the sake of discussion, we classify the
terminal knot, K2.5, as the radio ``hot spot'' in the context of similar
X-ray detected features in other powerful jets (see below and
\S~\ref{section-hotspot}).  \citet{sti93} noted also an 8.5 GHz VLA detection
of this terminal feature by \citet{pat92} from a 100 sec.  observation,
though no image is presented.  These four datasets form the basis of our
radio analysis of the jet. At the two higher frequencies, these data are
supplemented by the only other available VLA datasets -- sparse snapshots
from the more compact configurations to help fill in the shorter ($u,v$)
spacings (see Table~\ref{table-1} for a summary of the data). 

The data were obtained from the NRAO archive and calibrated in AIPS
\citep{bri94} utilizing scans of the calibrators 3C~48 and 3C~286 to set the
amplitude scale.  In the high-resolution A-configuration 15 GHz experiment
(program AB414; Table~\ref{table-1}), 3C~48 is heavily resolved so could not
be relied on for this purpose.  Fortunately, the phase calibrator 1803+784
was monitored frequently at this frequency by the UMRAO \citep{all03} and was
found to be fairly stable over the two month period bracketing the VLA epoch
(within $\sim$7$\%$ of the average of 3 Jy over 7 observations), so the
amplitude scale was set with this data. The calibrated ($u,v$) data were
exported into the Caltech DIFMAP package \citep{she94} for self-calibration
and imaging.  The multi-frequency images were registered on the image peaks,
i.e., the arcsecond-scale radio core. This registration must be good to
within a small fraction of the VLA beams because the majority of the VLBI
flux is concentrated within a few 10's mas (\S~\ref{section-vlbi}). 

Improving on the previous studies, we were able to separate and detect
emission from the radio hot spot {\it and} the jet at all four frequencies
(Figure~\ref{figure-2}).  The 1.5 GHz image is a ``super-resolved'' image,
created by reconvolving the data with a beam 1/3rd of the uniformly weighted
one. An excess due to the jet is apparent (Figure~\ref{figure-2}). There is
also an apparent excess adjacent to the core, on the side opposite of the
visible jet; a similar excess is seen in the 5 GHz image. This could be
suggestive of real emission on the counterjet side, although it is equally
likely that they are spurious due to the gaps in ($u,v$) coverage of the
datasets. 

\begin{figure*}
\epsscale{1.9}
\plotone{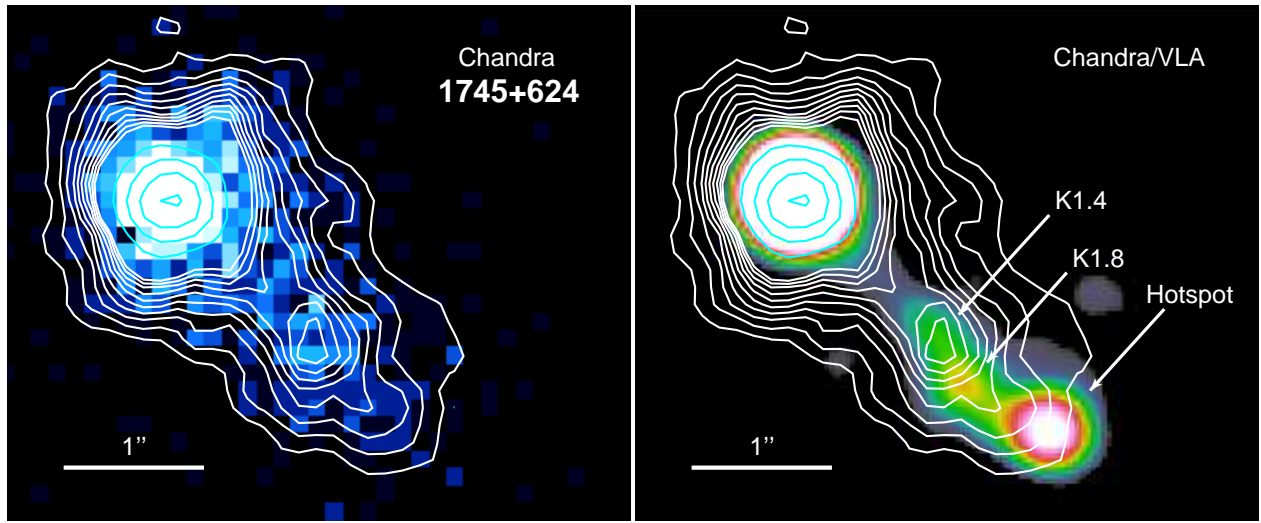}
\figcaption[f5.eps]{\label{figure-5}
Smoothed {\it Chandra} X-ray contours plotted over the full band X-ray   
[left] and VLA 8.5 GHz [right] images (plotted logarithmically).
The X-ray image was
rebinned to 1/4th the native ACIS pixel size of 0.492\arcsec\ and the
outer contours are spaced by 0.5 counts/pixel from 0.5 to 5 counts/pixel.
The main X-ray detected radio features (cf. Figure~\ref{figure-4}) are   
labeled.} \end{figure*}

We considered the individual knot spectra (\S~\ref{section-spectra}) but were
concerned with the differing ($u,v$) coverages affecting the spectral
measurements of the jet. To mitigate against this, we tapered and convolved
the three higher-frequency images with a common 0.35\arcsec\ beam (see e.g.,
Figures~\ref{figure-3}~\&~\ref{figure-5}) and integrate the radio jet
emission over several beamwidths when we measured the "total" jet spectral
index between the 3 frequencies (\S~\ref{section-spectra}). 

Additional polarization leakage-term calibration of the 5 GHz dataset was
performed using observations of the calibrator, 1748--253, from an
adjacent program AM337. The electric vector position angle (EVPA) of the
resultant polarization image was determined by setting the EVPA of 3C~48
to 106$\deg$. The integrated rotation measure (RM) of 28.4$\pm$0.5 radians
m$^{-2}$ \citep{ore95} toward 1745+624 amounts to a small correction of
--6\deg\ at 5 GHz, and this correction has been applied to the
polarization map in Figure~\ref{figure-3}. 

At 5 GHz, the initial segment of the radio jet is highly polarized at
$\sim$20$\%$ out to knot K1.4 with stronger interknot polarization of
30$\%$. The RM corrected map shows the EVPA perpendicular to the direction
of the jet along most of this length, i.e., the implied magnetic field is
parallel to the jet -- quite typical in powerful one-sided jets
\citep{bri84}. Also typical is the low polarization in the core
($\sim$4$\%$). 

The parallel-to-the-jet magnetic field configuration roughly continues
into K1.8 and the terminal radio feature, which we classify as a hot spot
based on its terminal position and compactness. The hot spot is unresolved
with an average deconvolved major axis of 0.15\arcsec\ and aspect ratio of
0.5 elongated at position angle of 254\deg\ (fits consistent between the
5, 8.5, and 15 GHz datasets). This outer portion of the jet exhibits lower
radio polarization levels of $\sim$5--6$\%$. The polarization pattern in
K1.8 is not uniform and field cancellation may account for its low
polarization. 

The dimensions of the jet knots, K1.4 and K1.8, are not obviously resolved
in our highest resolution (8.5 and 15 GHz) images though these are the
frequencies at which the jet is faintest. We set an upper limit to their
angular sizes (diameters) of 0.2\arcsec\ based on these data. 

Lastly, a new radio knot is found in the highest resolution (15 GHz) image
0.4\arcsec\ from the nucleus (K0.4). Though not well-separated from the
nucleus in the lower-resolution images, some excess emission presumably
from K0.4 is seen connecting the core to the outer jet in all three
lower-frequency maps (Figure~\ref{figure-2}). Its identification in the 15
GHz data helps delineate the overall (projected) path of the radio jet
from parsec-scales (out to $\simgt$0.1\arcsec;  \S~\ref{section-vlbi}) to
kiloparsec scales (\S~\ref{section-describe}).

\subsection{Hubble Space Telescope}

A deep Hubble Space Telescope (HST) image of 1745+624 was obtained from
the STScI archive to search for optical emission from the jet.  The data
analyzed was taken with STIS using the CCD in CLEAR (unfiltered) imaging
mode.  This image is sensitive over a broad range in the optical band
($\sim$2,000--10,000 \AA) with an effective wavelength of 5,740 \AA.  Four
roughly equal exposures totaling just under 3,000 sec.
(Table~\ref{table-1}) were aligned, then cosmic-ray rejected and combined
using the CRREJECT task in the STSDAS package in IRAF\footnote{IRAF is
distributed by the National Optical Astronomy Observatories, which are
operated by the Association of Universities for Research in Astronomy,
Inc., under cooperative agreement with the National Science Foundation.}. 

The quasar appears prominently in the image and has slightly saturated the
inner few CCD pixels (the following description follows
Figure~\ref{figure-4} closely). In the vicinity of the radio jet, a faint
resolved optical source is apparent just north of the jet (between K1.8
and the hot spot), but it is clearly unassociated with the extended radio
source. 

We modeled the central source with a series of elliptical isophotes using
the ELLIPSE task in IRAF, successfully out to $\sim$1.2\arcsec.  Other
than a clear source $\sim$0.95\arcsec\ roughly south of the quasar, this
subtraction revealed the alluded to optical emission apparently coincident
with K0.4.  As this is adjacent to the central mildly saturated source,
and similar features at two different position angles about the nucleus
are also seen in the subtracted image (one is lined up with the direction
of a diffraction spike), it is difficult to gauge its reality. It is
formally a 5$\sigma$ detection in a 0.15\arcsec\ radius circular aperture
(contains 79$\%$ of encircled energy) in the central source subtracted
image (Figure~\ref{figure-4}), where $\sigma$ here is the standard
deviation in the background in apertures adjacent to the jet. Some of the
optical emission may indeed originate from the radio source; its
radio-to-optical spectra index of $\sim$1 (Table~\ref{table-2}) is quite
typical in quasar jets \citep{sam04}, so we tentatively identify it as an
optical counterpart to the radio knot K0.4. This knot is unresolved from
the nucleus in the {\it Chandra} image so is not discussed in the
subsequent sections. 

We analyzed the outer knots and hot spot similarly with 0.2\arcsec\ radius
apertures (87$\%$ encircled energy) centered on the radio positions in the
unprocessed image and found no net positive signal.  Measurements of the
Poisson noise in these apertures result in a formal 3$\sigma$ point source
detection limit of 0.01 $\mu$Jy. However, experience with similarly
configured STIS CLEAR imaging data \citep{sam04} allow us to more
practically gauge detection limits for such jet knots -- we favor the more
conservative 3$\sigma$ limit of 0.06 $\mu$Jy derived from the previously
published work. 

An extinction correction of 10$\%$ was estimated using NED's quoted values
from the \citet{sch98} maps and though basically negligible, has been
applied to the optical measurements (Table~\ref{table-2}).

\begin{figure*}
\plotone{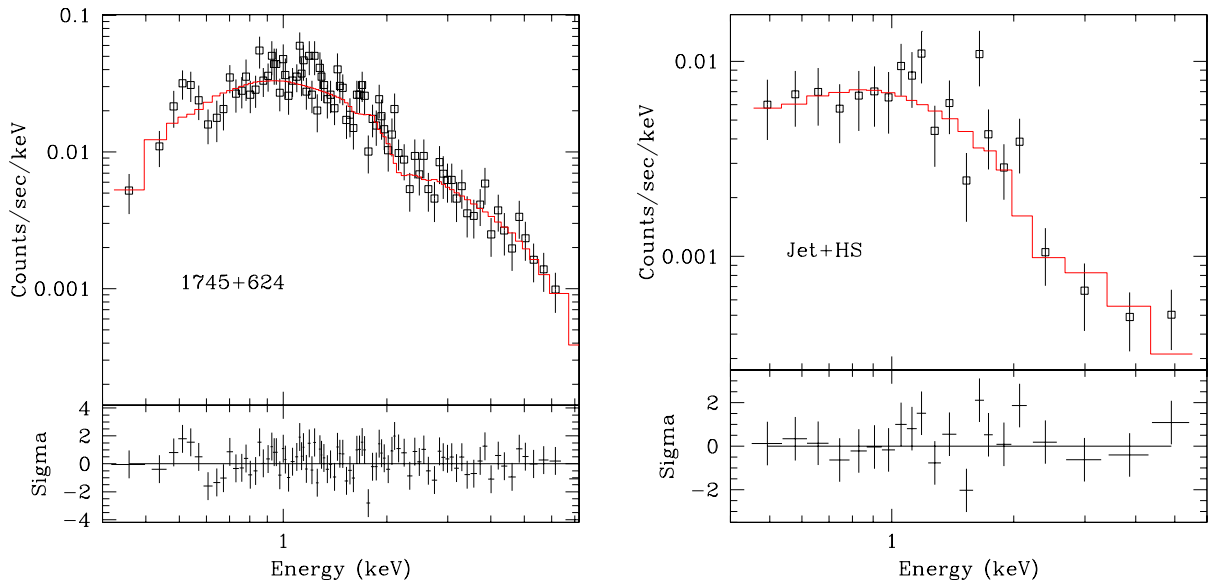}
\epsscale{1}
\figcaption[f6.eps]{\label{figure-6}
Top panels: Observed X-ray spectra of the core [left] and extended
jet+hot spot [right] emission from 1745+624.  The power-law plus   
absorption models are
drawn as lines over the data (squares with error bars). Both models
assumed a fixed local (galactic) absorption with an additional intrinsic  
($z$=3.89) absorbing component allowed in the core spectrum
(\S~\ref{section-core}). The bottom panels show the residuals from the
model.}
\end{figure*}

\subsection{{\it Chandra} X-ray Observatory\label{section-chandra}}

The quasar 1745+624 was observed with the ACIS camera aboard the {\it
Chandra} X-ray Observatory \citep{wei02} and we retrieved this data from
the {\it Chandra} archive. In this observation, the source was placed
30\arcsec\ from the default aim-point position on the ACIS-S backside
illuminated CCD, chip S3 (Proposers' Observatory Guide,
POG\footnote{http://cxc.harvard.edu/proposer/POG/index.html}). The data
were collected in a full readout mode of 6 CCDs resulting in 3.241 seconds
readout time.  The quasar's count rate of 0.056~cts~s$^{-1}$ gives 6-7$\%$
pileup in the core for this choice of the readout mode. After standard
filtering, the effective exposure time for this observation was 18,312~sec
(Table~\ref{table-1}). 

The X-ray data analysis was performed with the CIAO~3.2
software\footnote{http://cxc.harvard.edu/ciao/} using calibration files
from the CALDB3 database.  We ran {\tt acis\_process\_events} to remove
pixel randomization and to obtain the highest resolution image data. The
X-ray position of the quasar (R.A.= 17 46 14.057, Decl.= +62 26 54.79;
J2000.0 equinox) agrees with the radio position to better than
0.2$\arcsec$, which is smaller than {\it Chandra}'s 90$\%$ pointing
accuracy of 0.6\arcsec\ \citep{wei02}. The X-ray images were rebinned to
1/4 the native ACIS pixel-size of 0.492\arcsec\ in the final image; the
extended X-ray jet is already quite obvious in this image
(Figure~\ref{figure-5}). 

To determine the extent of the PSF at the location of the quasar, we ran a
ray-trace using the {\it Chandra} Ray Tracer
(CHART)\footnote{http://cxc.harvard.edu/chart/} and then
MARX\footnote{http://space.mit.edu/CXC/MARX/} to create a high S/N
simulation of a point source.  We modeled the quasar core as a point
source with the energy spectrum given by the spectral fitting described
below. For the simulation, we added 7$\%$ errors to account for
uncertainty in the ray-trace model \citep{sch00,jer04}.  We then extracted
a radial profile from both the {\it Chandra} data and the simulated point
source image assuming annuli separated by 0.45\arcsec\ and centered on the
quasar. The PSF was normalized to match the peak surface brightness of the
core. 

A comparison between the quasar profile and the simulated PSF shows an
excess of counts above the PSF beginning at $\sim 1.2$\arcsec\ from the
quasar centroid.  Therefore, in our analysis of the quasar's core X-ray
emission (\S~\ref{section-core}), we define a 1.2\arcsec\ radius circular
region centered on the X-ray centroid which excludes any overlap with the
region defined for the extended jet. For our analysis of the extended
emission (\S~\ref{section-xjet}), we defined two pie-shaped regions based
on the radio images -- one for the jet, and another for the terminal hot
spot.  Spectral analysis was performed in {\it Sherpa} \citep{fre01} using
counts in the energy range 0.3--7~keV.

\begin{figure*}
\epsscale{1.9}
\plotone{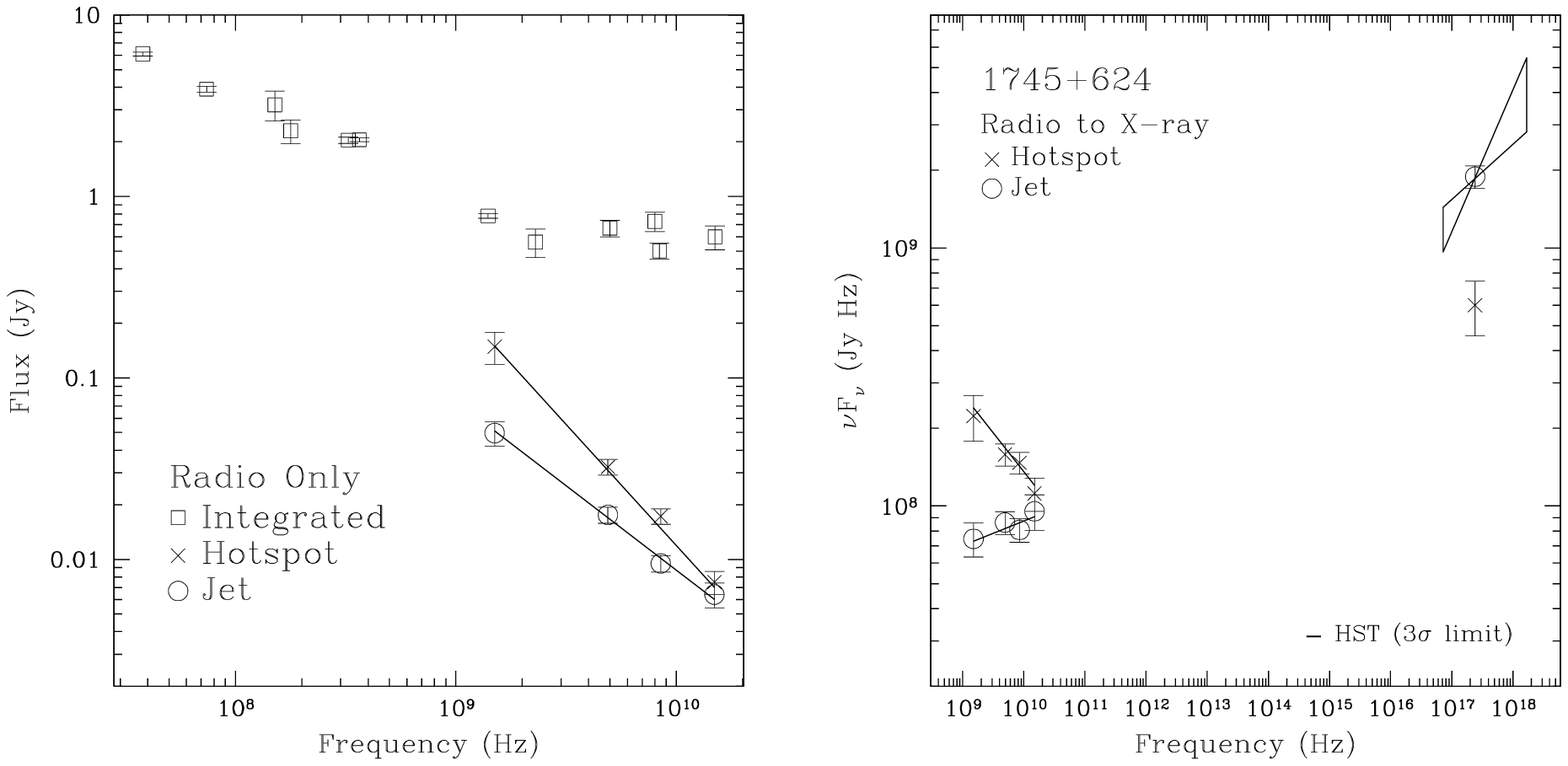}
\figcaption[f7.eps]{\label{figure-7}
[left] Integrated radio spectrum of 1745+624 along with the
hot spot and jet emission spectra from this work.
Lines are best fit power-laws to the hot spot and jet data. Integrated
measurements are at 38 \citep{hal95}, 74 \citep[VLSS;][]{coh06}, 151
\citep{hal93,vis95}, 178 \citep{gow67}, 325 \citep{ren97}, 365
\citep{dou96}, and 1400 MHz \citep{con98}, and 2.3, 5, 8 GHz
\citep{vol04}, 8.5 and 15 GHz (this work).
[right] Radio-to-X-ray spectral energy distributions of the total jet and
hot spot emissions in 1745+624. The bowtie around the X-ray jet point
represents the 1$\sigma$ measurement of the spectral slope (this is
suppressed in the case of the hot spot because its X-ray spectrum is not
well-constrained). }
\end{figure*}

\subsubsection{The Quasar Core X-ray Emission\label{section-core}}

Based on the PSF analysis described above, a 1.2\arcsec\ radius circular
region encircles about 91$\%$ of the total quasar counts. A total of 1,132
counts are detected in this region corresponding to 1,107.6$\pm 34.1$ net
counts.  We fit the spectrum between 0.3 and 7~keV with an absorbed power
law model assuming the photoelectric absorption model with abundance
tables from \citet{and89}.  Our analysis included the pileup model
specified by \citet{dav01}, which gives the fraction of piled photons
within 6--9$\%$.  The best-fit power law model has a photon index,
$\Gamma_{\rm X} = 1.85^{+0.07}_{-0.12}$ and a total absorbing Hydrogen
column of $N_{\rm H}=1.27^{+0.27}_{-0.25}\times 10^{21}$~cm$^{-2}$ for an
absorber at $z$=0 ($\chi^2=77.8/82$ dof, and we give 1$\sigma$ errors for
one significant parameter). Note that the fitted column is greater than
the equivalent column within the Galaxy (3.3$\times 10^{20}$~cm$^{-2}$) in
the direction of the quasar at high significance. This model gives a
0.5--2~keV flux of 1.5$\pm 0.2 \times 10^{-13}$~erg~cm$^{-2}$~s$^{-1}$
and 2--10~keV flux of 3.0$\pm 0.3\times
10^{-13}$~erg~cm$^{-2}$~s$^{-1}$ for the core. 

If we now assume that the quasar spectrum is absorbed by two absorption
components, a known (and fixed) Galactic column at $z$=0 and an intrinsic
absorber located at the quasar redshift, $z$=3.89, we obtain an intrinsic
$N_{\rm H}=2.35^{+0.78}_{-0.71}\times 10^{22}$~cm$^{-2}$ and the best-fit
photon index of $\Gamma_{\rm X}= 1.74^{+0.08}_{-0.14}$ ($\chi^2$=78.5/82
dof); see Figure~\ref{figure-6}.  This large intrinsic absorption column
confirms the previous {\it ASCA} result from analysis of the total
emission \citep{kub97} and is in agreement with the average column
observed in samples of high-redshift radio-loud quasars \citep{bas04}. 
These are comparable to the absorptions observed in broad absorption line
(BAL) quasars \citep{bro05}, where it is usually associated with an
outflow.  The photon index is also typical of that observed in lower
redshift quasars with X-ray jet detections \citep[e.g.,][]{gam03}.  In
this model, the 0.5--2~keV and 2--10~keV fluxes are equal to 1.6$\pm 0.2
\times 10^{-13}$~erg~cm$^{-2}$~s$^{-1}$ and 3.4$\pm 0.2 \times
10^{-13}$~erg~cm$^{-2}$~s$^{-1}$, respectively; these correspond to quasar
X-ray luminosities, $L_{\rm X}$(0.5--2~keV) = 1.6$\times
10^{46}$~erg~s$^{-1}$ and $L_{\rm X}$(2--10~keV)=5.7$\times
10^{46}$~erg~s$^{-1}$ (unabsorbed and K-corrected). 

We determined that there is no Fe-line present in the quasar spectrum with
a 3$\sigma$ upper limit on the equivalent width of 210 eV, consistent with
previous {\it ASCA} results \citep{kub97}. The {\it Chandra} data are also
consistent with little or no variability in both the X-ray flux and
spectrum since previous observations of (the total emission from) the
quasar \citep{fin93,kub97}.

\subsubsection{The Extended X-ray Emission\label{section-xjet}}

We extracted the spectrum of the total extended X-ray emission (from the
jet and hot spot) assuming a pie-shaped region between 1\arcsec\ to
2.75\arcsec\ centered on the quasar core. The assumed background region
was equal to a pie region spanning the same radii, but excluding the
jet+HS region. A total of 241 counts corresponding to 224.5$\pm 15.4$ net
counts was detected in this region. The binned spectrum was fit with an
absorbed power law model, and the best fit photon index for a fixed
Galactic column is equal to $\Gamma_{\rm X} = 1.71^{+0.13}_{-0.12}$
($\chi^2$=17.9/19 dof). This gives a 0.5--2~keV flux of 3.1$\pm 0.3 \times
10^{-14}$~erg~cm$^{-2}$~s$^{-1}$, and 2--10~keV flux of 6.1$\pm 0.6
\times 10^{-14}$~erg~cm$^{-2}$~s$^{-1}$ (Figure~\ref{figure-6}). 

Since we found that the radio hot spot spectrum is significantly steeper
than that of the jet (\S~\ref{section-spectra}), we attempted to analyze
the jet and hot spot emissions separately, assuming pie-shaped regions
between 1\arcsec\ to 2.05\arcsec\ and 2.05\arcsec\ to 2.75\arcsec,
respectively, centered on the quasar core.  The background region was
equal to the annulus with the same radii excluding the jet and hot spot
regions.  A total of 181 counts were detected in the jet region which
result in 173.2$\pm 13.7$ net counts. We binned the spectrum to have a
minimum of S/N=3 and fit an absorbed power law model to the jet spectrum
within the 0.3--7~keV energy range assuming the Galactic equivalent
Hydrogen column of 3.3$\times 10^{20}$~cm$^{-2}$. We obtain the best fit
photon index $\Gamma_{\rm X} = 1.62^{+0.16}_{-0.17}$ ($\chi^2$= 10.1/14
dof) corresponding to a 0.5-2~keV flux of 2.2$\pm 0.2 \times
10^{-14}$~erg~cm$^{-2}$~s$^{-1}$ and 2-10~keV flux of 5.1$\pm 0.5 \times
10^{-14}$~erg~cm$^{-2}$~s$^{-1}$. Similarly, for the hot spot (net
29$\pm$6.4 counts only), we found $\Gamma_{\rm X} = 2.1\pm0.6$ (1$\sigma$)
and fluxes 6.9$\times 10^{-15}$~erg~cm$^{-2}$~s$^{-1}$ (0.5--2~keV) and
7.9$\times 10^{-15}$~erg~cm$^{-2}$~s$^{-1}$ (2--10~keV). These results
suggest that, as in the radio band, the hot spot's X-ray spectrum is
steeper than the jet, although the formal difference is not well
determined because of the large uncertainty in the hot spot spectrum
($\Gamma^{\rm HS}_{\rm X}-\Gamma^{\rm Jet}_{\rm X}$=0.5$\pm$0.6).  The
model results from this paragraph are the ones used in the subsequent
discussion. 

We also allowed the total absorption column of the jet emission to vary
and obtained a slightly steeper photon index $\Gamma_{\rm X} =
1.74^{+0.36}_{-0.33}$ and a 3$\sigma$ upper limit to the absorption column
of $< 2 \times 10^{21}$~cm$^{-2}$.  This fitted photon index is consistent
within 1$\sigma$ of the value found with the column fixed. We thus conclude
that there is little gained by allowing the column to vary. 

In the {\it Chandra} image, the X-ray intensity ratio of the knots K1.4 to
K1.8 is $\sim$0.6:0.4. Assuming that the X-ray spectrum does not change
drastically between these two regions (there is no significant difference
in radio spectra; \S~\ref{section-spectra}), we estimated their fluxes
using their respective ratios of the total flux (Table~\ref{table-2}).

\subsection{Assessment of the Broad-band Jet Spectra\label{section-spectra}}

Taking our flux measurements of the hot spot at the four radio frequencies,
we fit a spectral index of $\alpha$=1.28$\pm$0.10
($F_\nu\propto\nu^{-\alpha}$), compared to $\alpha$=1.45 measured by
\citet{bec92} from mostly the same data (Tables~\ref{table-1}
\&~\ref{table-2}; Figure~\ref{figure-7}). In comparison to the individual jet
knot measurements (below), we believe this result is more robust because the
hot spot is bright, compact, and well-detected at the observed frequencies.
The fitted spectral index is insensitive to the presented low resolution 1.5
GHz measurement ($\alpha$=1.29$\pm$0.15 omitting this measurement in the
fit). This 1.5 GHz measurement though may be contaminated by emission from
the adjacent K1.8 (Figure~\ref{figure-2}). Since the flux ratio of K1.8 to HS
at the other bands is about 1:3, reducing the 1.5 GHz flux of the hot spot by
25$\%$ will change the spectrum to 1.20$\pm$0.09 which is still within the
uncertainty. 

The jet knots K1.4 and K1.8 are well-separated from the bright nucleus in the
5, 8.5 and 15 GHz images. These are the first reported detections of the {\it
jet} at frequencies other than 5 GHz \citep{bec92,pat92}. We measured a
"total" jet spectral index of 0.95$\pm$0.18 between these 3 frequencies on
the maps reconvolved to a common 0.35\arcsec\ beam (\S~\ref{section-radio}). 
Spectral indices measured from the full-resolution images give larger
uncertainties (Table~\ref{table-2}) because the signal-to-noise ratio of the
individual jet knot detections are low.  Including the ``super-resolved'' 1.5
GHz image, the resultant 4-frequency radio spectral index is 0.93$\pm$0.09,
which we adopt as the best determined spectral index of the jet
(Table~\ref{table-2}; Figure~\ref{figure-7}). 

The overall impression is that the jet knot radio spectra are fairly steep
-- on the steep end of what is observed in more nearby examples
\citep{bri84}.  As indicated by previous work \citep{bec92}, the hot spot
radio spectrum is significantly steeper than the jet ($\alpha_{\rm
HS}$-$\alpha_{\rm jet}$=0.35) and this is also suggested to be the case in
the X-ray data (\S~\ref{section-xjet}).  The slope of the hot spot
spectrum is as steep as the kpc-scale feature detected in GB~1508+5714
\citep{che05,che06} and is similar to those measured in comparably
high-redshift radio galaxies \citep[e.g.,][and references therein]{deb01}. 
These steep radio spectra are probably due to a combination of the
increased importance of IC/CMB losses on the higher energy electron
population and higher $(1+z)$ rest-frame frequencies sampled by the
observations -- this effect may be important when comparing low and
high-redshift jets (\S~\ref{section-z}). 

In addition to the radio spectra, the optical upper limits of all the
extended features preclude a simple power-law extrapolation of the radio
data into the X-ray band (i.e., $\alpha_{\rm r}$ and $\alpha_{\rm ro} >
\alpha_{\rm rx}$; Figure~\ref{figure-7}). This behavior is common among
the X-ray detected jets in quasars \citep[e.g.,][]{sam04}. One should bear
in mind that this assumes that the radio-to-X-ray radiation all originate
in the same emitting region, and also, the optical limits assumed a point
source.  Though the radio knots are unresolved at $\sim$0.15--0.2\arcsec\
resolution (\S~\ref{section-radio}), it does not preclude the optical
emission being more extended or that there are multiple emitting
components. 

Although there are complicating factors in the measurements of the radio
and X-ray spectra as discussed above, our analysis suggests that the
power-law slopes measured in the two bands are different for the jet.
Formally, this difference, $\alpha_{\rm r}-\alpha_{\rm x}$=0.31$\pm$0.19,
is significant. No similar evidence is evident in the hot spot spectra, as
we are limited by the low statistics of its X-ray detection.

\section{Discussion\label{section-discuss}}

At $z$=3.89, the 1745+624 system displays a fully formed jet as observed
by us when the Universe was only about 12$\%$ of its present age.  For its
projected length $L = 18$ kpc, and assuming a constant hot spot advance 
velocity $\beta_{\rm adv} \sim 0.1-1$ \citep[see in this context][but also 
comments at the end of \S~\ref{section-hotspot}]{sch95}, the lifetime of 
the radio source is roughly $t_{\rm life} \sim L/ (\beta_{\rm adv} c \, 
\sin \theta) \sim$ Myr (the jet viewing
angle $\theta \sim 10\deg$ is taken here for illustration; see below). 
This structure signals long-term active accretion at a very early cosmic
epoch, along with a roughly stable black hole/accretion disk axis
\citep[e.g.,][]{beg84}.  Such large-scale jets (deprojected $\sim$100 kpc
in this case) at high-redshifts are not necessarily rare, though they have
not been systematically searched for and studied. \citet[][and manuscript
in preparation]{che05} has only recently identified $\sim$10 other jets on
a comparable scale in a VLA study of a sample of $z>$3.4 (out to $z$=4.7)
flat-spectrum-core radio sources; 1745+624 remains the most
radio-luminous example observed (\S~\ref{section-intro}). 

Despite the extreme luminosity of the jet in 1745+624, its extended radio
and X-ray emissions appear quite similar to other more local examples in
regard to its morphology, spectral energy distribution, and radio
polarization.  This is quite surprising as higher-$z$ jets have propagated
through the (supposedly different) intergalactic medium of the early
Universe \citep[e.g.,][]{ree89}.
In the following, we first consider available evidence for relativistic
beaming in the 1745+624 jet on both parsec and kpc-scales
(\S~\ref{section-describe}).  Then, we discuss the emission properties of
the kpc-scale jet (\S~\ref{section-discussjet}) and the hot spot
(\S~\ref{section-hotspot}), confronting the multiwavelength data with
expectations from different emission models. Along the way, we make
comparisons to other X-ray jets at low and high redshift. Finally, the
broad-band data allow us to estimate the energetics of the jet, which we
compare to the accretion parameters of the quasar nucleus
(\S~\ref{section-energetic}).

\subsection{Radio Constraints on Relativistic Beaming 
of the Jet\label{section-describe}}

Several lines of evidence suggest that relativistic beaming in the
1745+624 jet is significant on the observed scales.  Both the radio (VLBI
and VLA-scale) and X-ray structures appear one-sided to our detection
limits.  The best constraints on the sidedness comes from the presented
2.3 GHz USNO map (jet to counter-jet ratio $>$50; \S~\ref{section-vlbi})
and proper motion studies with VLBI. Twelve epochs of higher resolution 8
GHz data from the same USNO program, obtained over $\sim$2.5 years
\citep{pin04}, show a basically stationary inner ($\sim$1--1.5 mas
distant) component, although velocities of a few $c$ are allowed by the
formal fit (B.G. Piner, 2005, private communication).  Maps from the
Caltech-Jodrell Bank survey obtained over a longer time baseline at 5 GHz
reveal three superluminal components presumably further down the jet
\citep[][this is the highest redshift object in the sample -- see Figure~1
in the latter]{tay94,ver03}. Converting to our adopted cosmology, their
proper motions correspond to apparent speeds ranging from $\sim$6$c$ up to
$\sim$16$c$. This restricts the maximum angle of the small-scale jet to
our line of sight to be $<$7--19\deg. 

The superluminal VLBI jet is oriented at PA$\sim$205--208\deg\ in the sky
\citep{tay94}, increasing gradually to $\sim$216\deg\ out to $\sim$110 mas
(Figure~\ref{figure-1}; \S~\ref{section-vlbi}). This further increases out
to the visible arcsecond-scale structure (the hot spot PA=227\deg).  The
maximal (projected) changes in position angle between the different
features is 6\deg\ (cf. Table~\ref{table-2}). The required intrinsic bend
in the jet to account for these changing PAs is $\sim$6\deg\ $\times
\sin(20\deg)$$\sim$2\deg, or even less (supposing the jet inclination
$<$20\deg\ as indicated by the proper motions). 

Since we can trace the VLBI-scale emission out to the observed {\it
Chandra} scales, the VLBI proper motions (and only small intrinsic bends)
constrain the kiloparsec-scale jet to be aligned within $\simlt$25\deg\ to
our line-of-sight, with a likely range of $\sim$10--20\deg, or less. 
Radio asymmetry studies of large scale quasar jets require jets like this
to be relativistic \citep[$\Gamma$$>$1, with strict upper limits of
$\Gamma$$<$2--3;][]{war97}; these constraints have bearing on the
following discussion of the origin of the X-ray emission in the 1745+624
jet. 

\subsection{X-ray Emission Mechanisms in the Large-scale
Jet\label{section-discussjet}}

The absence of optical emission and the steep radio spectra of the jet
knots in 1745+624 preclude a straightforward power-law extrapolation of
the low energy (synchrotron) emission into the X-ray band
(\S~\ref{section-spectra}). The resultant convex broad-band spectral
energy distribution (SED; Figure~\ref{figure-7}) is a common
characteristic of X-ray emitting knots in the jets of powerful quasars
\citep[e.g.,][]{schwartz00,sam04,mar05,har06}. As widely discussed,
(unbeamed) inverse Compton emission also can not account for the observed
excess of X-rays, and it is evident that this applies also to the present
example at very high-redshift (\S~\ref{section-basic}). We explore the
hypothesis that the bright X-ray emission is produced by {\it
relativistically beamed} inverse-Compton scattering of the CMB photons
\citep{tav00,cel01} as advocated by many workers. The main feature of this
model is that in order to fulfill energy equipartition, significant
beaming of the large-scale quasar jet emission is required
(\S~\ref{section-beamed}).  ``Non-standard'' synchrotron models have also
been proposed to reproduce the overall convex radio-to-X-ray SEDs and we
discuss this possibility in \S~\ref{section-synchrotron}. The main
difference between these models is the expected diverging
redshift-dependence of the X-ray emission, and we discuss issues and
prospects for distinguishing such a dependence from the observations
(\S~\ref{section-z}).

\subsubsection{Basic Considerations\label{section-basic}}

Generally, one can calculate the magnetic field in a synchrotron 
emitting radio source if we assume magnetic field energy equipartition 
with the radiating particles (corresponding to the minimum total 
energy of the system). In the 1745+624 radio jet, we use the observed 
radio properties of K1.4 and K1.8 (assuming negligible
contribution from relativistic protons and no beaming, $\delta$=1) to
calculate fairly high values of $B_{\rm eq, \, \delta=1} \approx 180$
$\mu$G and $200$ $\mu$G, respectively.  These values are more than an
order of magnitude higher (due to the exceptionally luminous nature of
this radio jet; \S~\ref{section-intro}) than those computed in
lower-redshift quasar jets \citep{kat05}, but are comparable to values
inferred for lower power FRI radio galaxies \citep{sta06}. The
corresponding magnetic field energy densities of the knots, $U_{\rm B} =
B_{\rm eq, \, \delta=1}^2/ 8 \pi \approx 1.3-1.6 \times 10^{-9}$
erg\,cm$^{-3}$, are $\sim$6--7 times greater than the CMB energy density
at the quasar's redshift, $U_{\rm CMB} = 4 \times 10^{-13} \, (1+z)^4$
erg\,cm$^{-3}$ $\approx 2.3 \times 10^{-10}$ erg\,cm$^{-3}$. The latter
quantity is, however, comparable to the energy densities of the knots'
synchrotron photons (in the case of sub-relativistic jet velocity),

\begin{equation} U_{\rm syn} \approx {L_{\rm syn} \over 4 \pi \, R^2 \, c} , 
\label{equation-usyn}
\end{equation}

\noindent namely, $U_{\rm syn} \approx$ 2.8 and 3.8 $\times 10^{-10}$
erg\,cm$^{-3}$ for K1.4 and K1.8, respectively. In this estimate, we
approximated the knots as spheres with radii, $R = 0.2\arcsec / 2 \approx
0.7$ kpc (\S~\ref{section-radio}), and took the synchrotron luminosities,
$L_{\rm syn} \approx 10 \times L_{\rm 5 \, GHz} \approx 5-7 \times
10^{44}$ erg\,s$^{-1}$, as appropriate for the radio continuum extending
over a few decades in frequency with spectral index $\alpha_{\rm r}
\approx 1$.  For the spectral index $\alpha_{\rm r}
\approx 1$, the expected X-ray flux densities ($F_{\rm X}$) from the
synchrotron self-Compton (SSC) plus IC/CMB processes are simply: 

\begin{equation} F_{\rm X}^{\rm ssc} + F_{\rm X}^{\rm IC/CMB} \approx
\left({\nu_{\rm r} \over \nu_{\rm X}}\right) \, \left({U_{\rm syn} +
U_{\rm CMB} \over U_{\rm B}}\right) \, F_{\rm r}.
\label{equation-ssc}\end{equation}

\noindent Using the 5 GHz radio flux measurements ($F_{\rm r}$), this
severely underpredicts ($\sim$0.054 nJy (K1.4) and $\sim$0.074 nJy (K1.8)
at 1 keV) the observed X-ray jet emission by factors of 87 and 42 in knots
K1.4 and K1.8, respectively (Table~\ref{table-2}).

\subsubsection{Relativistic Beamed X-ray Emission?\label{section-beamed}}

The bulk velocity of kpc-scale quasar jets are likely to be at least
mildly relativistic \citep[$\Gamma > 1$, ][]{bri94b,war97}. This can
significantly influence the expected inverse Compton X-ray fluxes
calculated above. Specifically, relativistic beaming ($\delta > 1$)
increases the IC/CMB emission, and decreases the SSC one.  In 1745+624,
the superluminal VLBI jet can be traced out to kpc-scales
(\S~\ref{section-describe}), so the beamed IC/CMB X-ray emission will
dominate over that produced by SSC in the visible jet. Assuming energy
equipartition as an additional constraint, the expected IC/CMB flux is
(for $\alpha\approx$1): 

\begin{equation} F_{\rm X}^{\rm IC/CMB} \approx \left({\delta \over
\Gamma}\right)^2 \left({\nu_{\rm r} \over \nu_{\rm X}}\right) 
\, \left({\Gamma^2 \, U_{\rm CMB} \over \delta^{-10/7} \, U_{\rm B}}\right) 
\, F_{\rm r} \, \, \, \propto \, \, \, \delta^{24/7}, \label{equation-iccmb}
\end{equation}

\noindent since in the jet rest frame (denoted by primes) $U'_{\rm CMB}
\approx \Gamma^2 \, U_{\rm CMB}$, and the equipartition magnetic field
$B'_{\rm eq} = B_{\rm eq, \, \delta=1} \, \delta^{-5/7}$. 
Here, $\Gamma$ is the jet bulk Lorentz factor, and
$U_B = B_{\rm eq, \, \delta=1}^2 / 8 \pi$. This gives the required $\delta
= 4.6$ and 3.9 for the knots K1.4 and K1.8, respectively, to explain the
observed X-rays via inverse Compton emission.  In this interpretation, the
(small) decrease of the required Doppler factor along the jet is a direct
consequence of the fading X-rays with the corresponding brightening of the
radio emission (Figure~\ref{figure-5}; cf. similar multi-wavelength
morphologies in other well-studied jets like, 3C~273 \citep{mar01,sam01},
0827+243 \citep{jor04}, PKS~1127--145 \citep{sie02}, and PKS~1136--135
\citep{sam06}). 

Although there is evidence that the 1745+624 jet is highly relativistic on
parsec scales (\S~\ref{section-describe}), there are few independent
constraints (the one-sidedness of the radio and X-ray jets) to directly
support the fairly high Doppler factors on these large scales, as implied in
this interpretation. The inferred Doppler factors require minimum Lorentz
factors, $\Gamma\geq (\delta + \delta^{-1})/2$ = 2.4 (K1.4) and 2.1 (K1.8).
These approach the strict upper limits on the speeds of large-scale quasar
jets deduced from radio asymmetry studies \citep[$\Gamma$$<$2--3;][]{war97}. 
If we assume the 1745+624 jet flow is moving near these maximal values, the
derived Doppler factors require that the jet is inclined within
$\simlt$12--15\deg\ to our line of sight, in agreement with the values
inferred in \S~\ref{section-describe}.  Given the discussed Lorentz factors
of $\sim$2--3, the Doppler factor is fairly insensitive to jet angles
considered; for instance, $\delta$=5.5--3.6 ($\Gamma$=3) and
$\delta$=4.0--3.2 ($\Gamma$=2.1) for jet angles, $\theta$=0\deg--15\deg. Thus
we can {\it just} plausibly explain the level of X-ray jet emission in this
jet via the IC/CMB mechanism if we push the speeds to be consistent with
upper limits inferred from previous studies, together with the small viewing
angles.  It should be stressed that these estimates assumed a single-zone
emitting region, and there are generally more free parameters in both Compton
and synchrotron models invoking substructure in the jet
\citep[e.g.,][]{cel01,sta02,sie06}. 

Related to the beaming issue, we should remark that the \emph{observed}
radio power of the 1745+624 jet (excluding the hot spot region) is very
large, $L_{\rm 5 \, GHz} \approx 10^{44}$ erg\,s$^{-1}$, suggesting its
synchrotron luminosity of order $\sim 10^{45}$ erg\,s$^{-1}$ or greater.
Such enormous radio powers from \emph{large-scale jets} (exclusive from
extended radio lobes) are quite extreme, so invoking even moderate beaming
factors would reduce the intrinsic jet frame luminosity (by 1--2 orders of
magnitude) to more comfortable levels. 

This relativistically beamed IC/CMB model also requires that there is a
sufficient number density of low energy electrons to upscatter CMB photons
into the observed {\it Chandra} band.  Radio emission would be radiated by
these electrons at very low-frequencies (10's MHz in this case) and
subarcsecond-resolution radio imaging capabilities at these energies are not
widely available.  Curiously, \citet{sti93} noticed that the {\it integrated}
radio spectrum of 1745+624 contains a steep-spectrum excess at
low-frequencies; this quasar would be catalogued as a steep-spectrum radio
quasar in low-frequency surveys and/or if it were local (the $1+z$ shift in
observed frequencies). In our own compilation of radio data from the
literature (Figure~\ref{figure-7}), we find $\alpha$=0.55$\pm$0.01 from 38
MHz to 1.4 GHz and $\alpha$$\sim$0 at the higher frequencies where our
imaging observations were taken. If we extrapolate the observed individual
spectra of the (dominant $\alpha$$\sim$0) core and steep-spectrum extended
component to lower frequencies, it implies that the latter dominates the
total low-frequency emission. The well-studied $z$=0.158 quasar 3C~273 shows
a similar low-frequency excess apparent down to $\sim$10's MHz \citep{cou98}
and resolved low-frequency imaging shows its extended X-ray emitting
\citep[e.g.,][]{mar01,sam01,jes06} radio jet does indeed dominate the total flux at
low frequencies \citep[cf. figure~A1 in][]{con93}.  Further, although this
low-frequency emission in 1745+624 is probably dominated by the (brighter and
steeper spectrum at cm-wavelengths;  \S~\ref{section-hotspot}) hot spot, this
does not preclude a low-frequency contribution from the jet. Interestingly,
the X-ray spectral index of the jet (Table~\ref{table-2}) matches the
low-frequency integrated radio one, as would be expected in the IC/CMB
scenario.  Future low-frequency imaging observations can clarify this by
showing the relative contributions of the jet and hot spot in this and 
similar systems \citep[see][for future prospects]{harris06}. 

\subsubsection{Synchrotron Models for the X-ray 
Emission\label{section-synchrotron}}

In lieu of the beamed IC/CMB interpretation, synchrotron models with
non-standard and/or multiple electron components are able to account for
the ``excess'' X-ray emission \citep{der02,sta02,sta04,jor04}. 
These models are attractive because in nearby low-power FRI jets, it is
widely believed that synchrotron X-ray emission is produced
\citep[e.g.,][]{har06}, although they do not show the same convex SEDs,
and emit at much smaller (few kpc) scales than the discussed quasar jets
(10's--100's kpc scales). The challenge remains to explain the strong
upturn in the X-ray band (i.e., the small values of $\alpha_{\rm ox}$) 
observed in powerful jets such as in 1745+624. 

While it is true that an increased energy density of the CMB photons at $z
> 1$ will result in a stronger radiative cooling of the ultra-relativistic
electrons, the maximum electron energy available assuming efficient
(although realistic) continuous acceleration process is still high enough
to allow for production of an appreciable level of keV synchrotron
photons.  In the case of the 1745+624 jet, the comoving energy density of
the magnetic field (by assumption close to equipartition with the
radiating electrons, as given above) is lower than the comoving energy
density of the CMB photons, $U'_{\rm B} < U'_{\rm cmb}$, if only
$\delta^{5/7} \Gamma > 2.5$. Thus, assuming even very moderate bulk
velocity and beaming, the jet electrons cool mainly by
inverse-Comptonization of the CMB photon field.  In the equipartition
derived magnetic field of $B$=200 $\mu{\rm G}$, the electrons emitting the
highest energy photons detected ($\sim$6 keV, observed) have electron
energies, $\gamma = E_{\rm e} / m_{\rm e} c^2 = (\nu_{\rm Hz} (1+z) /
\delta B_{\rm \mu{\rm G}})^{1/2}\sim 10^8$. Since their cooling are
dominated by inverse Compton losses, the appropriate timescale can be
evaluated roughly as $t_{\rm cool}' = 3\,m_e\,c / 4 \sigma_{\rm T} \gamma
U_{\rm CMB}' \sim 4.4$ yrs. This evaluation is rough, because in fact the
considered high energy electrons are expected to radiate in a transition
between Thomson and Klein-Nishina cooling regimes, since

\begin{equation}
\left({\Gamma \, \varepsilon_{\rm cmb} \over m_{\rm e} c^2}\right) \,
\left({E_{\rm keV} \over m_{\rm e} c^2}\right) \sim 1 \, ,
\end{equation}

\noindent with the energy of the CMB photon $\varepsilon_{\rm cmb} \approx
(1+z)$ milli-eV, and the jet bulk Lorentz factor $\Gamma$ of the order of
a few \citep[see in this context][]{der02}. For such parameters and the
considered electron energies, the `optimistic' electron acceleration
timescale $t'_{\rm acc} \sim \zeta \, r_{\rm g} / c \sim 0.01$ yr, where
$r_{\rm g}$ is the electron gyroradius and $\zeta \sim 10$ is the
efficiency factor depending on the jet plasma conditions
\citep[see][]{aha02}. This is still much shorter than the timescale for
radiative losses, while the latter is order of magnitudes longer than the
timescale for the electron escape from the emission region.

The postulated stochastic acceleration mechanism can produce a
flat-spectrum high-energy electron component, with the total energy
density not exceeding the magnetic field energy density \citep[see][for a
discussion]{sta02}. The \emph{observed} synchrotron spectrum of such
particles (from the unresolved emitting region) is expected to show a
spectral index modified (increased) by the radiative losses. With the
condition of continuous injection of freshly accelerated electrons
(characterized by a power-law energy spectrum $n'(\gamma) \propto
\gamma^{-p}$), such a spectral index would be $\alpha = 0.5$ for any
electron index $p < 2$ \citep[see][]{hea87}, which is very close to the
observed X-ray spectral index ($\alpha_{\rm X} \approx
0.62^{+0.16}_{-0.17}$) of the 1745+624 jet. Moreover, the total luminosity
of such a high-energy synchrotron component, if indeed limited only by the
energy equipartition requirement, should be comparable to or less than the
luminosity of the low-energy (radio) synchrotron component. That is in fact
the case, since the \emph{total} $1$ keV luminosity of the 1745+624 jet
(excluding the terminal feature discussed in \S~\ref{section-hotspot}) is
$\sim 3 \times 10^{45}$ erg s$^{-1}$ while, as we argue in
\S~\ref{section-basic} above, the radio jet luminosity is $> 10^{45}$ erg
s$^{-1}$.  Note, that in the framework of this synchrotron interpretation,
the X-ray-to-radio jet luminosity ratio is not expected to change
systematically with redshift in contrast to expectations in an IC/CMB
interpretation, and this issue is discussed in the next subsection. 

\subsubsection{Redshift Dependence of the X-ray Emission?\label{section-z}}

The simplest versions of synchrotron and IC/CMB models give significantly
different predictions for the X-ray jet emission at high redshift because
of the strong increase of the CMB energy density $\propto (1+z)^4$
\citep[e.g.,][]{sch02}.  With all else equal, they predict divergences in
the observed X-ray to radio monochromatic luminosity
($f_{\nu}=\nu\,F_{\nu}$) ratio, $f_{\rm x}$/$f_{\rm r}$ exceeding a factor
of 100 or more at $z$$>$2 (cf. Equation~\ref{equation-iccmb}). This
difference should be readily apparent in observations of $z\sim$4 jets
like in 1745+624 if the X-rays were dominated by IC/CMB emission. X-ray
jets detected so far have observed $f_{\rm x}$/$f_{\rm r}$ values which
range over 4 orders of magnitude and there is no obvious sign of a
$(1+z)$-dependence in the latest large compilation of \citet{kat05}. 

However, there are complicating factors which will skew such correlations,
in both models, such as source-to-source (and knot-to-knot within a single
jet) variations in jet magnetic fields, electron energy densities, and
speeds.  Particularly in current samples biased toward ``blazars'', the
X-ray jet emission is particularly sensitive to the (very small) jet
angles to our line of sight if there is significant relativistic beaming
on these large scales.  For an IC/CMB origin, \citet[][cf. figure~4
therein]{che04} outlined such a scenario to account for the large $f_{\rm
X}/ f_{\rm r}$ ratio of the $z$=4.3 GB~1508+5714 quasar jet ($>$ 100;
among the largest value found thus far) in comparison to other known,
lower redshift ($z\simlt$2) jets due mostly to its extreme
redshift\footnote{$f_{\rm X}/ f_{\rm r}$ can be equivalently expressed as
the radio-to-X-ray power-law slope; $\alpha_{\rm rx}$=0.73 for
GB~1508+5714 \citep{che04} while $\alpha_{\rm rx} \sim 1.1-0.8$,
translating to $f_{\rm X}/ f_{\rm r} \sim 0.1-30$ at lower redshifts
\citep[see figures 5 and 9
in][respectively]{sam04,kat05}.\label{footnote-x}}.  This scenario can be
extended to the $z$=3.63 core-dominated quasar J2219--2719 where 6 total
extended X-ray counts were recently detected in an $\sim$8 ksec exposure
\citep{lop06}, and can be attributed to a $\sim$1 mJy radio feature
2\arcsec\ south of the quasar \citep[thus $f_{\rm X}/ f_{\rm r}>100$ also;
][and manuscript in preparation]{che05}. Two quasars showing particularly
low $f_{\rm x} / f_{\rm r}$ ratios, despite their fairly high-redshifts
($z$=1.4 and 2) also stood out -- these are especially lobe-dominated
quasars in comparison to the other examples, and a connection to beaming
was suggested \citep[see][for a discussion]{che04}. 

One caveat to note is that the Doppler factors required for the knots in
the 1745+624 jet (in the minimum-energy + IC/CMB framework) are rather
moderate, $\delta=4-5$, and similar to that found in the GB~1508+5714 case
\citep[][\S~\ref{section-intro}]{che04}. The required $\delta$ in these
two $z\sim$4 quasar jets are at the low end of the corresponding values
calculated for lower-redshift objects ($\delta\sim$4--25). In fact, these
few critical data points at high-$z$ help to define the trend discussed by
\citet[figure 10 therein]{kat05} that the maximal values of $\delta$
required to explain the X-ray jet emission as IC/CMB {\it decreases with
redshift}. Also, in contrast to the GB~1508+5714 and J2219--2719 cases,
the $f_{\rm x}$/$f_{\rm r}$ values of the 1745+624 jet are closer to those
observed at lower-redshifts (Table~\ref{table-2} and
footnote~\ref{footnote-x}).  Although second-order effects from the
intrinsic jet properties should be factored in, the low implied jet
Doppler factors and lack of a very obvious redshift-dependence of $f_{\rm
x}$/$f_{\rm r}$, can be taken as evidence against an IC/CMB
interpretation. In particular, if there were more extreme Doppler factors
than in the $z$$\sim$4 quasars observed so far, the large-scale jets would
be orders of magnitude brighter and will outshine the active cores in
X-rays via the IC/CMB process \citep{sch02}.  Currently however, no such
sources have been found in a growing number of high-$z$ radio-loud quasars
imaged with subarcsecond-resolution by {\it Chandra} \citep{bas04,lop06},
so where are the very highly beamed high-$z$ X-ray jets? There is of
course a natural Malmquist bias inherent in studying very high-redshift
objects, so this and other selection biases must be addressed in future
samples. 

On the other hand, if these trends persist with further observations, it
may instead reflect intrinsic differences between large-scale quasar jets
located at different redshifts and/or their different environments, which
are then probed with observations. For example, in the IC/CMB
interpretation, the trend may imply that high-$z$ jets of this kind are
intrinsically slower than their lower-redshift counterparts. This may be
due to a more disturbed environment (evidenced by the ``alignment effect''
of high-$z$ radio galaxies; e.g., Rees 1989) through which high-$z$ jets
are propagating. Such an scenario, in fact, was suggested early on to
explain the distorted morphologies observed in a large samples of high-$z$
($\sim$1.5--3) quasars and radio galaxies \citep[e.g.,][]{bar88}. 

As current X-ray studies have tended toward known well-studied radio jets
\citep[e.g.,][]{sam04,mar05} that are relatively local, it is important to
test if this and other trends persist and can be clarified with larger,
more homogenous samples of distant ($z>$2--4) jets.  In analogy to VLBI
proper motion studies, the earliest superluminal motions detected tended
to be the highest \citep{ver94} and larger ensembles of source
measurements now tend toward lower values \citep{ver03,kel04}. This may
mean that beaming factors of kpc-scale X-ray jets as inferred from IC/CMB
calculations are more typically smaller than the highest ones found thus
far.

\subsection{The Terminal ``Hot Spot'' \label{section-hotspot}}

Recent {\it Chandra} studies have aimed to distinguish the X-ray detected
terminal jet features (i.e., the hot spots) from the jets
\citep{hard04,kat05,tav05}. Many of these knots show much the same problem
as in the jet, i.e. the inability to extrapolate the radio-to-optical
power-law slopes smoothly in the X-ray band, and the underprediction of
not-beamed inverse-Compton (SSC and IC/CMB) emission to the observed. In this
context, it is useful to discuss the terminal feature K2.5 of the 1745+624
jet as a terminal hot spot.

Knot K2.5 follows the conventional empirical definition of a hot spot only
loosely \citep{bri94b}, as applying these strict criteria to such
high-redshift, small angular-size jets is quite restrictive.  Hot spots
are strong terminal shocks formed at the tips of powerful jets where the
strong compression of the jet magnetic field leads to a transverse
configuration and relatively high radio polarization. This expectation is
in good agreement with observations of sources located at low redshifts.
K2.5 is situated at the edge of the radio source and it is indeed compact
with a spatial extent of $\sim 0.15\arcsec \times 0.075\arcsec \approx 1
\times 0.5$ kpc (\S~\ref{section-radio}), which is typical for the known
X-ray detected hot spots in quasars and FR II radio galaxies.  However,
the parallel configuration of the magnetic field with respect to the jet
axis, as well its low degree of linear polarization relative to the rest
of the jet (Figure~\ref{figure-3}), are not expected, nor typical of the
hot spots resolved in more nearby radio sources \citep{bri84}. {\it This
may betray the dominance of an underlying jet flow in this feature.}

The 5 GHz flux of the hot spot in 1745+624 implies a monochromatic
luminosity, $L_{\rm 5 \, GHz} \approx 2.42 \times 10^{44}$ erg\,s$^{-1}$.
This is higher, although still comparable to the 5 GHz luminosities of hot
spots at lower redshifts \citep[e.g.,][]{kat05}.  Assuming energy
equipartition as in the jet, and a cylindrical geometry for the emitting
region with radius $0.15\arcsec/2 \approx 0.5$ kpc
(\S~\ref{section-radio}) and similar length, we calculate the hot spot
magnetic field, $B_{\rm eq, \, \delta=1} \approx 660$ $\mu$G. This also is
in rough agreement (though again slightly higher) with the minimum energy
magnetic fields derived for the other knots \citep{kat05}. 

The equipartition magnetic energy density of the hot spot, $U_{\rm B}
\approx 1.7 \times 10^{-8}$ erg\,cm$^{-3}$, is more than one order of
magnitude greater than the energy density of the CMB radiation at the
quasar's redshift. However, the energy density of the synchrotron photons
thereby, $U_{\rm syn} \approx 4.7 \times 10^{-9}$ erg\,cm$^{-3}$, is
roughly comparable to $U_{\rm B}$. In this evaluation, we took the hot
spot radio spectral index, $\alpha_{\rm r} \approx 1$, its synchrotron
luminosity $L_{\rm syn} \approx 10 \times L_{\rm 5 \, GHz}$, the geometry
as described above, and neglected relativistic corrections. The implied 1
keV flux of the hot spot due to the SSC plus IC/CMB processes is then
$\sim$0.18 nJy (Equation~\ref{equation-ssc}). As in the jet emission
(\S~\ref{section-basic}), this is inconsistent with the observed X-ray
flux, though here the deviation is less drastic (factor of $\sim$14;
Table~\ref{table-2}). However, one should keep in mind many of the
underlying assumptions in such calculations before claiming a true
disagreement between model predictions and the observations. In
particular, the hot spot structure is in fact unresolved, hence a slightly
smaller volume than considered, together with, e.g. small deviations from
energy equipartition and spectral curvature (see below), could foreseeably
bring the measurement into better agreement with the model prediction. 

However, if we take the observed excess in X-rays over the estimated SSC
flux at face value, this would not be the first such case for a hot spot.
\citet{hard04} has argued for a synchrotron origin of this X-ray excess in
many local hot spot sources, though in the case of 1745+624, the optical
upper limit precludes a straightforward extrapolation of the (very steep)
radio spectrum up to the X-ray band. On the other hand, \citet{tav05} has
argued that there is an IC/CMB contribution from an underlying
relativistic portion of the jet terminus (unresolved by our observations) to 
the few X-ray detected hot spots studied in several $z\sim 1$ quasars. 

The latter possibility is quite attractive in the case of 1745+624: as we
noted earlier, the terminal jet feature in this source does not have
typical hot spot-like radio polarization properties, and there could very
well be an unresolved portion of the jet mixed in.  If the multi-band
radiative output of K2.5 is dominated by a relativistic jet flow, this
would imply $\delta$=5 for an IC/CMB origin of the X-ray flux (using
Equation~\ref{equation-iccmb} with all the parameters discussed above).
This beaming factor is quite high, though comparable to values derived in
lower-redshift quasars \citep{tav05}. It is also slightly higher than the
values obtained for the proper jet (knots K1.4 and K1.8; see
\S~\ref{section-beamed}), though again, there are the usual uncertainties
in these calculations to keep in mind.  Regardless, the problem of
terminal X-ray/radio knots in this and many quasar jets can be viewed as a
simple extension of that in the jet knots. 

One relevant issue in this discussion is the apparent lack of a hot spot on
the side opposite of the nucleus to the prominent jet. If the advance speed
of the jet is sub-relativistic, then some emission should have been visible
on the counter-jet side. This can be reconciled by the fact that the usually
inferred (sub-relativistic) advance speeds of hot spots in powerful radio
sources assume that the jet thrust is balanced with the ram-pressure of the
ambient material within a typical galaxy group/poor cluster environment
\citep{beg84}.  However, if the jets propagate into the radio cocoon formed
in a previous epoch of radio activity, i.e. in an environment with a
substantially decreased number density and pressure, the advance velocity of
the jet may be close to $c$.  In such a case, combined Doppler and
light-travel effects may result in an apparent lack of a visible counter hot
spot. Such a scenario was proposed by \citet{staw04} for the 3C~273 jet.
Since the observed properties of 1745+624 and 3C~273 radio quasars are so
similar, one could also apply this argument to explain the lack of emission
visible on the counter-jet side in 1745+624. 

Let us finally discuss yet another issue regarding this dominant radio
feature, namely the extrapolation of its radio spectrum to lower
frequencies. As discussed in \S~\ref{section-beamed}, an extrapolation of
its observed $\sim$$1-10$ GHz continuum $\propto \nu^{-1.3}$ to lower
frequencies exceeds the extrapolation of the flat-spectrum core component
$\propto \nu^0$ at about $\sim 500$ MHz, and joins smoothly with the
observed integrated low frequency emission at $\sim 200$ MHz (see
Figure~\ref{figure-7} [left]). In other words, one can attribute the
$\sim$30--300 MHz integrated emission to the extended radio structure.
Since the extended emission is probably dominated by the brightest
feature, the hot spot (we can not exclude a contribution from the fainter
inner jet; \S~\ref{section-beamed}), this would imply a broken power-law
character of its synchrotron spectrum, $\propto \nu^{-0.5}$ for $\nu <
300$ MHz and $\propto \nu^{-1.3}$ for $\nu > 300$ MHz. In the standard
continuous-injection model of terminal features in powerful jets
\citep{hea87}, the spectral break would correspond to the energies,
$E_{\rm e} / m_{\rm e} c^2 \sim 10^3$. This low-frequency break is
expected in face of the large computed hot spot equipartition magnetic
field, $B_{\rm eq, \, \delta=1} \approx 660$ $\mu$G \citep[see a
discussion in][]{bru03,cheung05}.

\subsection{The Jet Energetics \label{section-energetic}}

The multiwavelength emissions of the 1745+624 jet allow us to estimate the
kinetic power of the outflowing plasma in this source. The kinetic power
of the ultrarelativistic electrons and magnetic field characterizing the
proper jet, under the minimum-power hypothesis, is roughly $L_{\rm j} \sim
L_{\rm B} + L_{\rm e} \sim 2 \times \pi R^2 \, c \Gamma^2 \, U'_{\rm B}$,
where $R \approx 0.7$ kpc is the jet radius and $U'_B = 1.5 \times 10^{-9}
\, \delta^{-10/7}$ erg s$^{-1}$ is the jet comoving magnetic field energy
density, as discussed in \S~\ref{section-discussjet}. Assuming for
illustration, a jet viewing angle $\sim 10\deg$ (see
\S~\ref{section-describe}) and jet bulk Lorentz factor at the kpc-scale
$\Gamma \sim 3$ (leading to the `comfortable' value of the jet Doppler
factor $\delta \sim 4.7$), one obtains a kinematic factor $\delta^{-10/7}
\, \Gamma^2 \sim 1$, and therefore a jet kinetic luminosity (regarding
solely ultrarelativistic jet electrons and the jet magnetic field) $L_{\rm
j} \sim 10^{45}$ erg s$^{-1}$. 

If we regard the terminal feature as a ``true'' hot spot, we can estimate
the kinetic energy of the jet. We approximate the broad-band synchrotron
spectrum of the hot spot by a broken power-law: $\propto \nu^{-0.5}$
between the assumed minimum synchrotron frequency $\nu_{\rm min} = 10$ MHz
and the break frequency $\nu_{\rm br} = 300$ MHz, followed by $\propto
\nu^{-1.3}$ between $\nu_{\rm br}$ and the assumed maximum synchrotron
frequency $\nu_{\rm max} = 10^{12}$ Hz (see \S~\ref{section-hotspot}). The
radiative efficiency factor, i.e. the ratio of the power emitted by the
hot spot electrons via synchrotron radiation, and the total power injected
to these electrons at the terminal shock, is: 

\begin{equation}
\eta_{\rm rad} \sim {\ln (\nu_{\rm max} / \nu_{\rm br}) \over \ln 
(\nu_{\rm max} / \nu_{\rm min})} \sim 0.7 \, .
\end{equation}

\noindent The reconstructed radio continuum of the hot spot implies also
its total synchrotron luminosity $L_{\rm syn} \approx 10 \times L_{\rm 5
\, GHz} \sim 2.5 \times 10^{45}$ erg s$^{-1}$. Thus, the total kinetic
power of the jet transported from the active center to the jet terminal
point is roughly:

\begin{equation}
L_{\rm kin} \sim {L_{\rm syn} \over \eta_{\rm rad} \, \eta_{\rm e}} \, ,
\end{equation}

\noindent where $\eta_{\rm e}$ is the fraction of the jet kinetic energy
transformed in the terminal shock to the internal energy of
ultrarelativistic (i.e.  synchrotron emitting) electrons. In the case of
energy equipartition between the electrons and the magnetic field reached
at the hot spot, one should expect very roughly $\eta_{\rm e} \sim 0.5$
\citep[see][for a wider discussion]{sik01}. This gives $L_{\rm kin} \sim
0.75 \times 10^{46}$ erg s$^{-1}$. In fact, $\eta_{\rm e} = 0.5$ should
be regarded as an upper limit, keeping in mind a possible proton 
contribution and work done by the outflow in pushing out the ambient 
medium, making $10^{46}$ erg s$^{-1}$ a safe lower limit for the
kinetic power of the 1745+624 jet. The jet power estimated at the 
beginning of this subsection is an order of magnitude lower than its 
total kinetic power constrained from the hot spot emission. This may 
indicate a dynamical role of non-radiating jet particles \citep[either 
cold electrons or protons; see in this context,][]{sik00}. 

Let us compare the kinetic power of the 1745+624 jet estimated above with
the accretion parameters of the quasar core. Due to the large distance of
the quasar, and hence related observational difficulties, these accretion
parameters cannot be evaluated precisely thus should be treated with some 
caution. Nevertheless, we note \citet{sti93} and
\citet{sk93} found a broad emission CIV line in the spectrum of the
1745+624 quasar at the observed wavelength $\lambda_{\rm CIV}^{\rm obs} =
7553$ \AA, and an observed ${\rm FWHM_{CIV}^{\rm obs}} = 64.9$ \AA. This
gives the intrinsic velocity dispersion: 

\begin{equation}
v_{\rm CIV} = c \, {{\rm FWHM_{CIV}} \over \lambda_{\rm CIV}} = c \, 
{{\rm FWHM_{CIV}^{\rm obs}} 
\over \lambda_{\rm CIV}^{\rm obs}} \approx 2576 \, {\rm km \, s^{-1}} \, .
\end{equation}

\noindent \citet{sti93} reported also a steep-spectrum ($\alpha_{\rm opt}
= 1.3$), non-variable and low-polarized optical continuum of the 1745+624
quasar, which we consider below as being produced by the circumnuclear gas
solely. In other words, we assume that the host galaxy and the nuclear
portion of the jet do not contribute significantly to the observed optical
emission of the discussed object. At the observed wavelength $6600$ \AA,
corresponding to the source rest frame wavelength $1350$ \AA, the detected
continuum flux is $F_{6600} \approx 0.08$ mJy. Thus, the intrinsic
continuum luminosity at the emitted frequency $1350$ \AA, is simply
$L_{1350} = 4 \pi d_{\rm L}^2 \, \left[\lambda F_{\lambda}\right]_{6600}
\approx 5.4 \times 10^{46}$ erg s$^{-1}$. These estimates allow us to find
the bolometric luminosity of the quasar core (assumed to be about ten
times the $V$-band core luminosity), $L_{\rm bol} \sim 10 \times L_{5500}
\sim 10 \, \left(\nu_{5500} / \nu_{1350}\right)^{1-\alpha_{\rm opt}} \,
L_{1350} \sim 8.2 \times 10^{47}$ erg s$^{-1}$, as well as the mass of the
central supermassive black hole via the virial relation
\citep{ves02,ves06}: 

\begin{equation}
{{\cal M}_{\rm BH} \over {\cal M}_{\odot}} = 5.4 \times 10^6 \, 
\left({v_{\rm CIV} \over 1000 \, {\rm km/s}}\right)^2 \, 
\left({L_{1350} \over 10^{44} \, {\rm erg/s}}\right)^{0.53} \simgt 1
\times 10^9.
\end{equation}

\noindent This can be taken as a lower limit in this case because the
relation was calibrated for the broad component of the CIV line, which was
not measured independently from the narrow one \citep{sti93,sk93};
including the latter underestimates the measured broad-line width
\citep{ves02} which could be $>$3,000 km s$^{-1}$.  The corresponding
Eddington luminosity is $L_{\rm Edd} \simgt 10^{47}$ erg s$^{-1}$ and the
Eddington ratio, $\Lambda \sim L_{\rm bol} / L_{\rm Edd} \sim 1-10$. The
latter is quite large and may indicate contribution of the emission due to
an unresolved portion of the jet to the nuclear optical continuum produced
by the accreting matter. 

Finally, the total observed $1.4$ GHz flux of the 1745+624 radio source
\citep[780.6 mJy; ][]{con98}, gives the total monochromatic radio
luminosity $L_{\rm R} \approx 1.6 \times 10^{45}$ erg s$^{-1}$ thus a
radio loudness parameter, ${\cal R} \sim 10^5 \, L_{\rm R} / L_{\rm V}
\sim 10^{3} - 10^{4}$. Note that the discussed object, which possesses the
most powerful observed radio jet known, is extremely `radio-loud', as the
standard division between radio-quiet and radio-loud sources \citep{kel89}
is ${\cal R} = 10$. With the obtained values of the Eddington ratio and
radio loudness, 1745+624 fits into the `radio-loud' trend formed on the
$\Lambda - {\cal R}$ plane by local ($z < 0.5$) FR I radio galaxies and
radio selected broad-line AGNs, as discussed in detail by \citet{sik06}.

\section{Summary\label{section-summary}}

We have analyzed multi-frequency radio, optical and X-ray imaging data for
the kpc-scale jet in the $z$=3.89 quasar 1745+624. This quasar hosts the
most powerful large-scale radio and X-ray jet yet observed, with
monochromatic radio and X-ray luminosities, $L_{\rm 5 \, GHz} \approx 1.3
\times 10^{44}$ erg s$^{-1}$ and $L_{\rm 1 \, keV} \approx 2.8 \times
10^{45}$ erg s$^{-1}$, respectively (excluding emission from the powerful
terminal ``hot spot''). Aside from its large radiative output, and related
large inferred equipartition magnetic field (in comparison to local
powerful jet sources), its properties (multi-wavelength morphology, radio
polarization, and broad-band spectral energy distribution) are broadly
similar to other lower-redshift quasar jets. This is unexpected in light
of the dramatic increase in the CMB energy density and strong cosmic
evolution of the intergalactic medium at such high-redshifts;  this should
have manifested in a dramatically different appearance of such large-scale
outflows. 

The spectral energy distributions of the resolved linear structures are
discussed.  In this jet, inverse-Compton scattered emission on the CMB
photons can just account for the X-ray emission if the jet is inclined
close to our line of sight ($\simlt$10\deg) and if it is also moving
relativistically, $\Gamma$=2--3. Several indirect arguments from the
multi-scale radio observations of 1745+624 support such values of the
large-scale jet kinematic parameters. The data are also consistent with a
synchrotron interpretation provided the electron acceleration timescale is
much shorter than that of the radiative losses, which we deem
likely. The main distinguishing feature of the two models should manifest
in drastically different X-ray/radio emission properties of such
high-redshift jets though with only three current $z>$3 examples, no
trends are yet apparent. 

Via the virial relation, we estimate that a ${\cal M}_{\rm BH} \simgt 1
\times 10^9 \, {\cal M}_{\odot}$ supermassive black hole resides at the
center of this galaxy.  The broad-band emission of the extended components
allow us to additionally estimate a jet kinetic power of $L_{\rm kin} \sim
10^{46}$ erg\,s$^{-1}$, which is a small fraction of the Eddington
luminosity ($L_{\rm kin}/L_{\rm Edd} \sim 0.1$) corresponding to this
black hole mass. As the bolometric luminosity of the quasar is $L_{\rm
bol} / L_{\rm Edd} \sim 10$, the inferred jet kinetic power seems to be
much less than the estimated accretion power. Despite this, the 1745+624
quasar, hosting the most powerful radio jet known, is extremely
`radio-loud' in comparison to more local radio selected quasars with a
radio-loudness parameter ${\cal R} \geq 10^{3}$, and is surprisingly
similar to local radio selected broad line AGNs regarding the accretion
parameters.

\acknowledgments
\begin{center}{\bf Acknowledgments}\end{center}

This work began while C.~C.~C. spent the initial year of his fellowship at
the MIT Kavli Institute. He is grateful to the High Energy Transmission
Grating group there for their hospitality, and the KIPAC at Stanford for
currently hosting his fellowship. \L .~S. acknowledges support by MEiN
through the research project 1-P03D-003-29 in years 2005-2008, and by the
ENIGMA Network through the grant HPRN-CT-2002-00321.  This research is
funded in part by NASA (A.~S.) through contract NAS8-39073 and through
Chandra Award Number GO5-6113X issued by the Chandra X-Ray Observatory
Center, which is operated by the Smithsonian Astrophysical Observatory for
and on behalf of NASA under contract NAS8-39073. 

The VLA is a facility of the National Radio Astronomy Observatory is
operated by Associated Universities, Inc. under a cooperative agreement
with the National Science Foundation. 
Based in part on observations made with the NASA/ESA Hubble Space
Telescope, obtained from the data archive at the STScI.  STScI is operated
by the Association of Universities for Research in Astronomy, Inc. under
NASA contract NAS 5-26555. 
This research has made use of the United States Naval Observatory (USNO) 
Radio Reference Frame Image Database (RRFID).
We thank Alan Fey for providing us these USNO datasets and Glenn Piner for
sharing results prior to publication. 
We thank Dan Harris for his practical comments on the discussion.

{}

\clearpage

\begin{table}
\begin{footnotesize}
\caption[]{\label{table-1} Multi-Telescope Archival Data for Quasar 1745+624}
\begin{center}
\begin{tabular}{lccccc}
\hline \hline
Instrument&Program&Date       &Frequency             & Exptime & Reference or Observer\\ 
(1)&(2)&(3)&(4)&(5)&(6)\\
\hline
VLA A-configuration & AB414 &  08-Sep-1991 &1.5                  & 500 & Becker et al. (1992)\\
VLA A-configuration & AB414 &  08-Sep-1991 &4.9                  & 500 & Becker et al. (1992)\\
VLA A-configuration & AP182 &  20-Feb-1990 &8.5                  & 100 & Patnaik et al. (1992)\\
VLA B-configuration & AM303 &  08-Sep-1990 &8.5                  & 140 & M. Malkan\\
VLA A-configuration & AB414 &  08-Sep-1991 &14.9                 & 810 & Becker et al. (1992)\\
VLA B-configuration & AM303 &  08-Sep-1990 &14.9                 & 200 & M. Malkan\\
VLA C-configuration & AM310 &  09-Dec-1990 &14.9                 & 200 & M. Malkan\\
HST STIS 50CCD-CLEAR& 8572  &  08-Sep-2000 &5.2$\times$10$^{5}$  &2,786& L. Storrie-Lombardi\\
{\it Chandra} ACIS-S3     & 4158  &  09-Sep-2003 &0.3--7 keV          &18,312& P. Strub \\
\hline \hline
\end{tabular}
\end{center}
(1) Telescope, and array configuration or camera-aperture.\\
(2) Observational program code.\\
(3) Date of observation.\\
(4) Frequency of observations in units of gigahertz unless indicated otherwise.\\
(5) Total integration time in seconds. At 8.5 and 14.9 GHz, the 
multi-configuration data were combined.\\
(6) The program principal investigator is listed unless we located a 
published reference to the data. 
\end{footnotesize}
\end{table}

\begin{table}
\begin{footnotesize}
\caption[]{\label{table-2} Multi-wavelength Flux Densities and Spectra of Extended 
Emission} 
\begin{center}
\begin{tabular}{lccccc}
\hline \hline
Region 		& K0.4		&K1.4	&K1.8	&K2.5 (hot spot)	&Total Jet*	\\
\hline
Distance (\arcsec) & 0.4		&1.4	&1.8	&2.5		&$\sim$0.8$-$2.0	\\
PA (deg.)	& 221	&224 &221 &227 	&225\\ 
\hline
$F_{\rm 1.5}$(mJy) 	& --		&--	&--	&148.7$\pm$29.7 &49.7$\pm$7.5\\
$F_{\rm 4.9}$(mJy) 	& --		&6.8$\pm$1.0	&9.3$\pm$1.4  &32.3$\pm$3.2 	&17.6$\pm$1.8 \\
$F_{\rm 8.5}$(mJy) 	& --		&2.9$\pm$0.4	&6.2$\pm$0.9  &17.3$\pm$1.7 	&9.5$\pm$1.0\\
$F_{\rm 14.9}$(mJy) 	& 3.4$\pm$0.7  	&3.0$\pm$0.6	&2.5$\pm$0.5  &7.5$\pm$1.1 	&6.4$\pm$1.0\\
$F_{\rm opt}$($\mu$Jy)	& 0.13(0.14)$\pm$0.03 &$<$0.06(0.06) &$<$0.06(0.06) &$<$0.06(0.06)&--\\
$F_{\rm 1~keV}$(nJy) 	& --		& 4.7$\pm$0.5 &3.1$\pm$0.4&2.5$\pm$0.6	&7.8$\pm$0.8\\
\hline
$\alpha_{\rm r}$& --            &0.96$\pm$0.33& 1.15$\pm$0.20& 1.28$\pm$0.10    &0.93$\pm$0.09 \\
$\alpha_{\rm ro}$&0.97(0.96)    & $>$1.01(1.01) &$>$1.03(1.03)  &$>$1.14(1.14)  & --\\
$\alpha_{\rm rx}$&--		&0.80	&0.84	&0.92		&0.83\\
$\alpha_{\rm x}$& --		&--	&--	&1.1$\pm$0.6		&0.62$^{+0.16}_{-0.17}$\\
\hline \hline 
\end{tabular}
\end{center}
Notes.---\\
--Knot names indicate their distances relative to the core in arcseconds.
The position angle (PA) is positive east of north defined to be zero degrees.\\

--The flux densities, $F_{\nu}$, are at the radio frequencies 1.5, 4.9,
8.5, and 14.9 GHz, optical (opt; 5.2$\times$10$^{5}$ GHz; extinction
corrected values are indicated in parentheses), and 1 keV X-rays. For the
hot spot and total jet, we converted the 2--10~keV fluxes from the
spectral fits into flux density using the fitted spectral indices
($\alpha_{\rm x}$) indicated. The flux densities of the individual
features come from the ratios of the (background subtracted) counts to the
total jet, K1.4:K1.8 = 0.6:0.4.\\

--The radio ($\alpha_{\rm r}$), radio-to-optical ($\alpha_{\rm ro}$),
radio-to-X-ray ($\alpha_{\rm rx}$), and X-ray ($\alpha_{\rm x}$;
$\Gamma_{\rm X}=1+\alpha_{\rm X}$) spectral indices where
$F_{\nu}\propto\nu^{-\alpha}$. The broad band spectral indices use 5 GHz
values (for K0.4, the 15 GHz measurement was converted to 5 GHz assuming
$\alpha_{\rm r}$=0.9).\\

*Jet fluxes in the radio were measured directly from the VLA maps. At 4.9,
8.5, and 14.9 GHz, the maps were reconvolved with a common 0.35\arcsec\
beam with ($u,v$) tapering applied at the two higher frequencies. At 1.5
GHz, we used the super-resolved map (Figure~\ref{figure-2}).\\
\end{footnotesize}
\end{table}

\end{document}